\documentclass[useAMS,usenatbib,usegraphicx]{mn2e}

\title[The Eridanus Supergroup]{Eridanus - A Supergroup in the Local Universe?}
\author[Brough et al.]
	{Sarah Brough$^{1}$,\thanks{E-mail: sbrough@astro.swin.edu.au}
	Duncan A.~Forbes$^{1}$, Virginia A.~Kilborn$^{1,2}$, Warrick
	Couch$^{3}$, \newauthor Matthew Colless$^{4}$
\\$^{1}$Centre for Astrophysics and Supercomputing, Swinburne University of Technology, Hawthorn, VIC 3122, Australia
\\$^2$Australia Telescope National Facility, CSIRO, P.O. Box 76, Epping, NSW 1710, Australia
\\$^{3}$School of Physics, The University of New South Wales, Sydney, NSW 2052, Australia
\\$^{4}$Anglo-Australian Observatory, P.O. Box 296, Epping, NSW 1710, Australia}

\begin{document}

\date{Accepted... Received...; in original form 2005}

\pagerange{\pageref{firstpage}--\pageref{lastpage}} \pubyear{2005}

\maketitle

\label{firstpage}

\begin{abstract}
We examine a possible supergroup in the direction of the Eridanus
constellation using 6dF Galaxy Survey second data release (6dFGS DR2)
positions and velocities together with 2MASS and HyperLEDA photometry.
We perform a friends-of-friends analysis to determine which galaxies
are associated with each substructure before examining the properties
of the constituent galaxies.  The overall structure is made up of
three individual groups that are 
likely to merge to form a cluster of mass $\sim7\times10^{13}M_\odot$.
We conclude that this structure is a supergroup.  We also examine the
colours, morphologies and luminosities of the galaxies in the region
with respect to their local projected surface density.  We find that
the colours of the galaxies redden with increasing density, the median luminosities are brighter with increasing environmental density
and the morphologies of the galaxies show a strong morphology-density
relation.  The colours and luminosities of the galaxies in the
supergroup are already similar to those of galaxies in clusters,
however the supergroup contains more late-type galaxies, consistent
with its lower projected surface density.  Due to the velocity
dispersion of the groups in the supergroup, which are lower than those
of clusters, we conclude that the properties of the constituent
galaxies are likely to be a result of merging or strangulation
processes in groups outlying this structure.
\end{abstract}


\section{Introduction}
\label{intro}
The paradigm of hierarchical structure formation leads us to expect
that clusters of galaxies are built up from the accretion and merger
of smaller structures like galaxy groups
(e.g. \citealt{blumenthal84}).  Although we observe clusters of
galaxies accreting galaxy group-like structures along filaments
(e.g. \citealt{kodama01,pimbblet02}) we lack clear examples of groups
merging together to form clusters -- a `supergroup'.  We define a
supergroup to be a
group of groups that will eventually merge to form a cluster.  The
first example of a supergroup was found recently at $z\sim0.4$ by
\cite{gonzalez05}.

Studies of the properties of galaxies suggest that they strongly
depend on the density of the environment in which they reside.
Studies of the properties of galaxies in clusters out to several
virial radii
(e.g. \citealt{dressler80,hashimoto98,balogh97,balogh98,kodama01,pimbblet02,wake05,tanaka05})
and in large galaxy surveys such as the Two-degree Field Galaxy
Redshift Survey (2dFGRS;
\citealt{lewis02,depropris03,balogh042df,croton05,hilton05}) 
and Sloan Digital Sky Survey (SDSS;
\citealt{blanton03,gomez03,baldry04,kauffmann04,tanaka04}) 
have determined that the properties of galaxies in the densest regions
differ from those in the least dense regions.  Those galaxies in the
densest regions are more likely to be early-type galaxies
(e.g. \citealt{dressler80,postman84,dressler97,tran01,lares04}),
redder
(e.g. \citealt{kodama01,pimbblet02,girardi03,blanton03,baldry04,tanaka04,wake05}),
with a lower star-forming fraction
(e.g. \citealt{balogh97,hashimoto98,zabludoff98,lewis02,gomez03,balogh042df,kauffmann04,wilman05}).
This suggests that star formation is being suppressed in denser
environments.  The projected surface density at which this difference
occurs is $\sim1-2$ galaxies Mpc$^{-2}$ \citep{lewis02,gomez03}.  This
is of the order of the density associated with poor groups of
galaxies.

However, it is still uncertain what the driving factor in the change
of properties is.  Whether it is a result of nature -- galaxies that
form in such environments have these properties, or nurture --
galaxies falling into the clusters are transformed by their
environment, is unclear.  The proposed mechanisms for this
transformation are: ram pressure stripping \citep{gunn72,quilis00},
strangulation
\citep{larson80,cole00}, harrassment \citep{moore96}, tidal interactions
\citep{byrd90,gnedin03} and galaxy mergers.  Of these, ram pressure 
stripping and harrassment are more likely to be dominant in the dense,
but rare, environments of clusters whereas mergers and strangulation
are more likely in the group environment where the velocity dispersion
of the group is similar to that of its constituent galaxies
\citep{barnes85,zabludoff98,hashimoto00}.  \cite{miles04} show that 
there is a dip in the luminosity function of low X-ray luminosity
groups ($L_X<10^{41.7}$ erg s$^{-1}$) consistent with rapid merging
currently occurring at these densities.

Whether nature or nurture is the dominating factor, the group
environment is clearly very important to the process of galaxy
evolution, not least because more than 50 per cent of galaxies in the
local Universe are found in groups \citep{eke04}.  
Given this dependence on environment and the importance of the
lower-density group environment, finding and studying supergroups is
vital to our understanding of how both clusters and their constituent
galaxies evolve.



The known concentration of galaxies in the region behind the Eridanus
constellation is a possible supergroup in the local Universe.  This
concentration was first noted by \cite{baker33}.
\cite{devaucouleurs75} found that his group 31, and groups associated
with NGC 1332 and NGC 1209 formed the `Eridanus Cloud'.  The Southern
Sky Redshift Survey (SSRS; \citealt{dacosta88}) determined that this
cloud lies on a filamentary structure with the Fornax cluster and
Dorado group to the south and in front of the `Great Wall'.

The Eridanus cloud lies at a distance of $\sim 21$ Mpc and includes
two optically classified groups of galaxies (NGC 1407;
\citealt{garcia93} and NGC 1332; \citealt{barton96}).  However, there
is some debate as to the nature of this region.  \cite{willmer89}
describe it as a {\it cluster} made up of three to four subclumps,
while \cite{omar05a} describe it as a {\it loose group} at an
evolutionary stage intermediate to that of the Ursa-Major and Fornax
clusters.
\cite{willmer89} calculated that the different subclumps are bound to
one another.

The NGC 1407 and NGC 1332 groups have previously been studied as part
of the Group Evolution Multiwavelength Study (GEMS;
\citealt{osmond04}).  This is an on-going study of 60 groups with the aim of
determining how groups, and their constituent galaxies, evolve.  The
groups were selected from optical catalogues and then compared with
the {\it ROSAT PSPC} pointings.  Groups were then selected such that
the group position was within $20^{\prime}$ of the {\it ROSAT}
pointing co-ordinates, the {\it ROSAT} exposure time was $>10,000$s
and the recessional velocity of the group was $1000 < v < 3000$ km
s$^{-1}$.  This ensured that there were enough photon counts to
confirm a detection and that the X-ray emission was neither so close
as to overfill the PSPC field of view nor too distant to be
detected. This resulted in a sample of 35 groups, to which a further
25 which had been previously studied with the PSPC
\citep{helsdon00,ponman96} were added.  Wide-field neutral hydrogen
(HI) observations of 16 GEMS groups in the Southern hemisphere were
made with the Parkes radiotelescope \citep{kilborn05}, these include
the NGC 1332 and NGC 1407 groups and their HI properties are discussed
later in this paper.

\begin{figure*}
\begin{center}

    \resizebox{30pc}{!}{
	\includegraphics{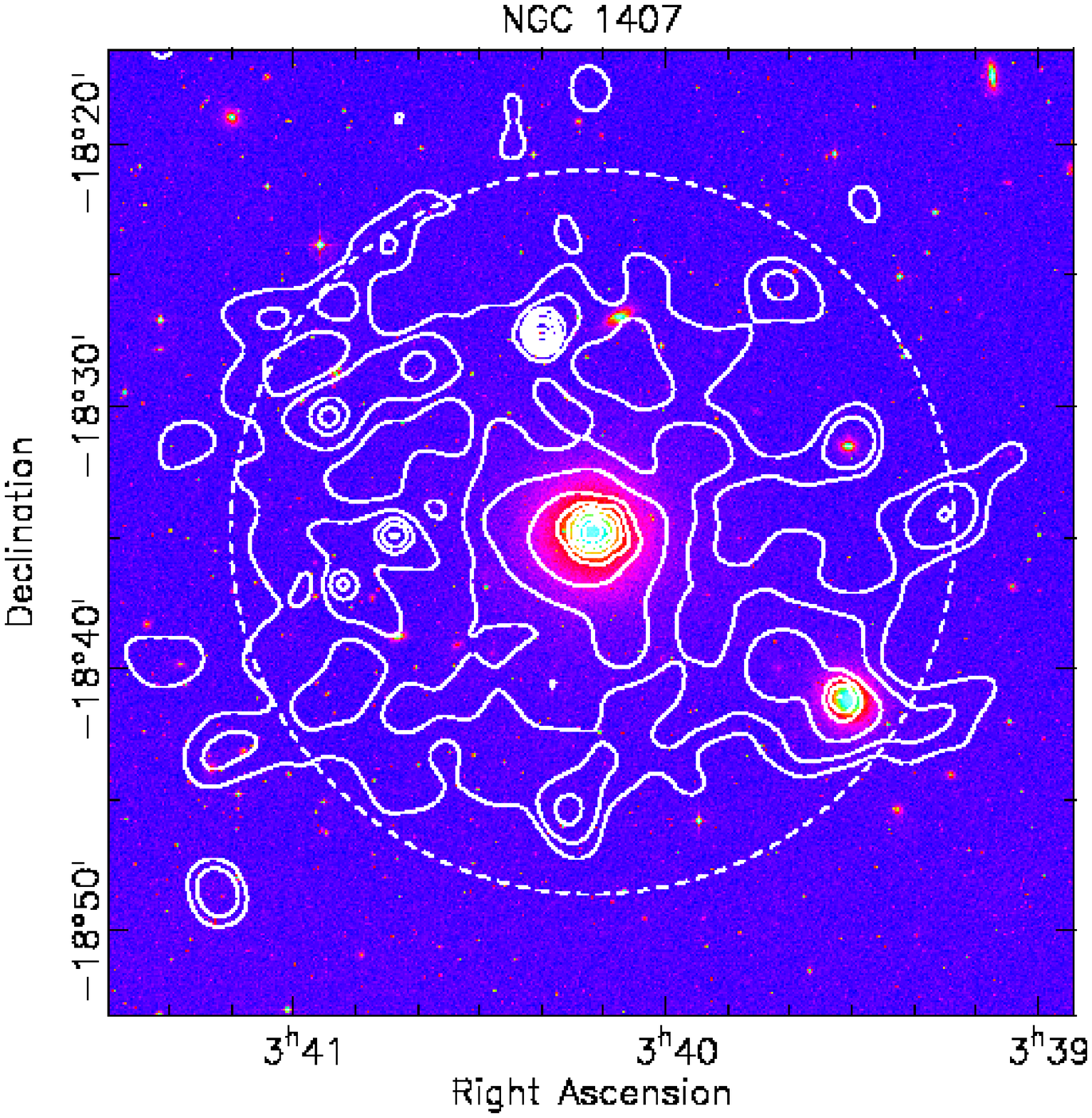}
	\includegraphics{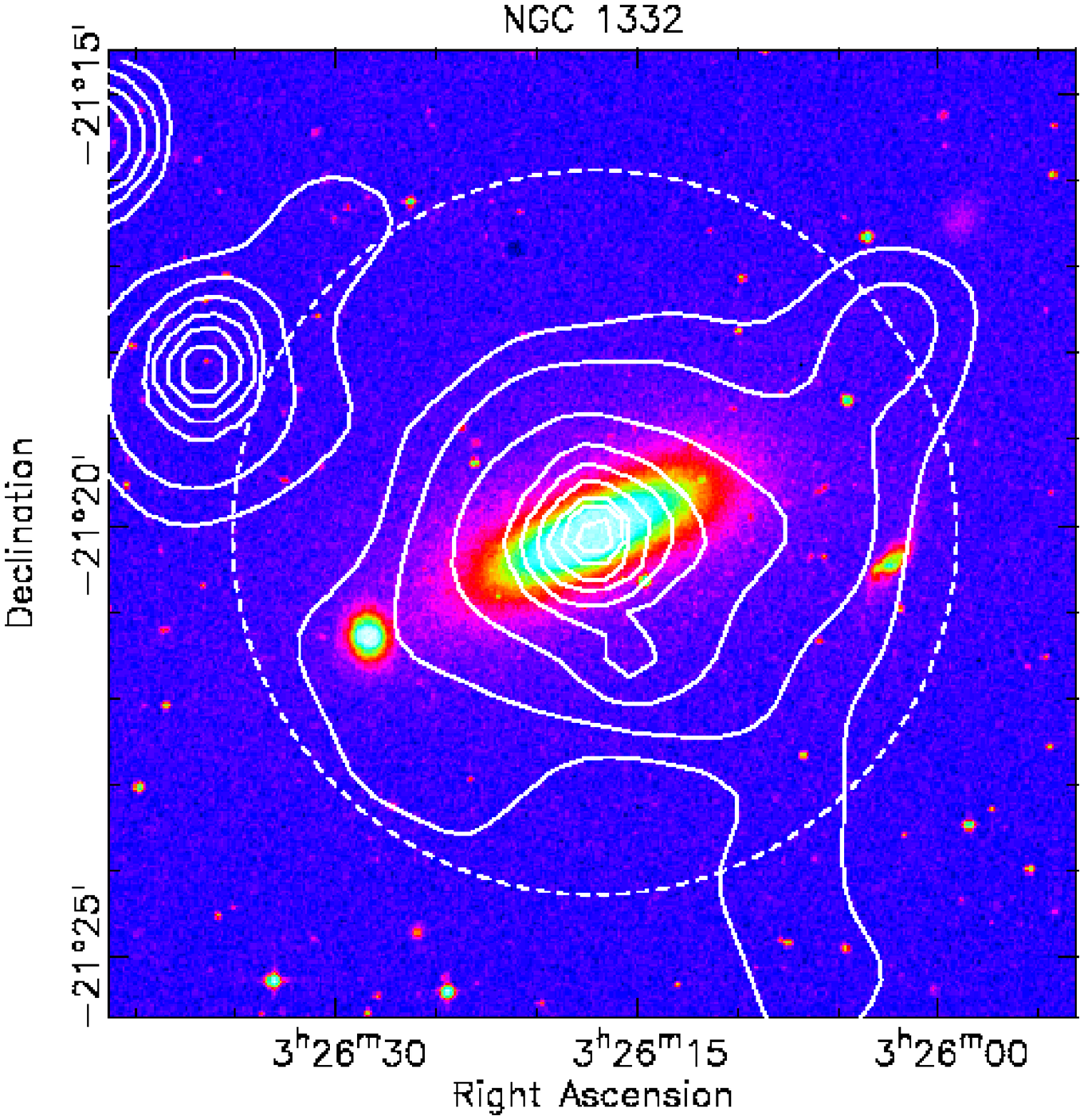}
    }
  \end{center}
\caption {{\it ROSAT PSPC} contours for NGC 1407 and NGC 1332 
overlaid on an optical DSS image from the GEMS archive (Osmond \&
Ponman, private communication).  The dashed circle represents the
radius ($r_{cut}$) at which the X-ray emission falls to the background
level.  The extended X-ray emission of NGC 1407 is clearly associated
with the group as a whole ($r_{cut}=105$ kpc).  In contrast, the more
compact X-ray emission seen around NGC 1332 is only associated with
NGC 1332 itself ($r_{cut}=28$ kpc).}
\label{x_ray_plots}
\end{figure*}

{\it ROSAT} data is therefore available for both the NGC 1407 and NGC
1332 groups from the GEMS archive (Osmond \& Ponman, private
communication).  Figure~\ref{x_ray_plots} shows that the X-ray
emission around NGC 1407 is symmetric, if irregular, and is emitted
from the intra-group gas, confirming the presence of a massive
structure.  In contrast, the X-ray emission from the NGC 1332 group is
associated with NGC 1332 itself, not with intra-group gas.
\cite{omar05a} suggest that there is intra-group gas associated
with NGC 1395, a large elliptical in this region, however no optically
selected group has previously been associated with this galaxy.

The advent of the 6dF Galaxy Survey (6dFGS; \citealt{jones04}) with
its public release of positions and velocities of galaxies in the
local Universe enables a reanalysis of this region to determine what
this structure is.  
Here we present new measurements allowing the first detailed dynamical
analysis of this region and its constituent galaxies.

\section{Data}
\subsection{6dFGS}

The 6dF Galaxy Survey (6dFGS; \citealt{jones04}) is a wide-area (the
entire Southern sky with $|b|>10^{o}$), primarily $K_s$-band selected
galaxy redshift survey.  The catalogue provides positions, recession
velocities, and spectra for the galaxies, along with $K_s$-band
magnitudes from the 2MASS extended source catalogue
\citep{jarrett00,jones04} and $r_F$- and $b_J$-band data from the
SuperCOSMOS catalogue
\citep{hambly01}.  However, at magnitudes brighter than $b_J\sim16$
mag the SuperCOSMOS data are unreliable owing to problems with
galaxy-galaxy deblending.  The second data release of the 6dFGS (DR2;
\citealt{jones05}) contains 71,627 unique galaxies.

Galaxy data were obtained from the 6dFGS DR2 for an area of radius 15
degrees ($95$ Mpc$^2$), centred on the position of NGC 1332, in the
velocity range 500 -- 2500 km s$^{-1}$.  The lower recession velocity
limit was chosen to avoid Galactic confusion.
This provides a sample of 135 unique galaxies, detailed in
Appendix~\ref{6df_galaxies}.  The 6dFGS database also provides 2MASS
$K_s$-band magnitudes where available.  We have used the 2MASS $K_s$
magnitudes within the 20th magnitude isophote (henceforth denoted as
$K$).  As 2MASS is $>99$ per cent complete to $m_K\sim13.1$
\citep{jarrett00} we assume that those galaxies without 2MASS data are
fainter than the 2MASS magnitude limit.

\subsection{NED}

\begin{figure}
\begin{center}

    \resizebox{30pc}{!}{ \rotatebox{38}{ \includegraphics{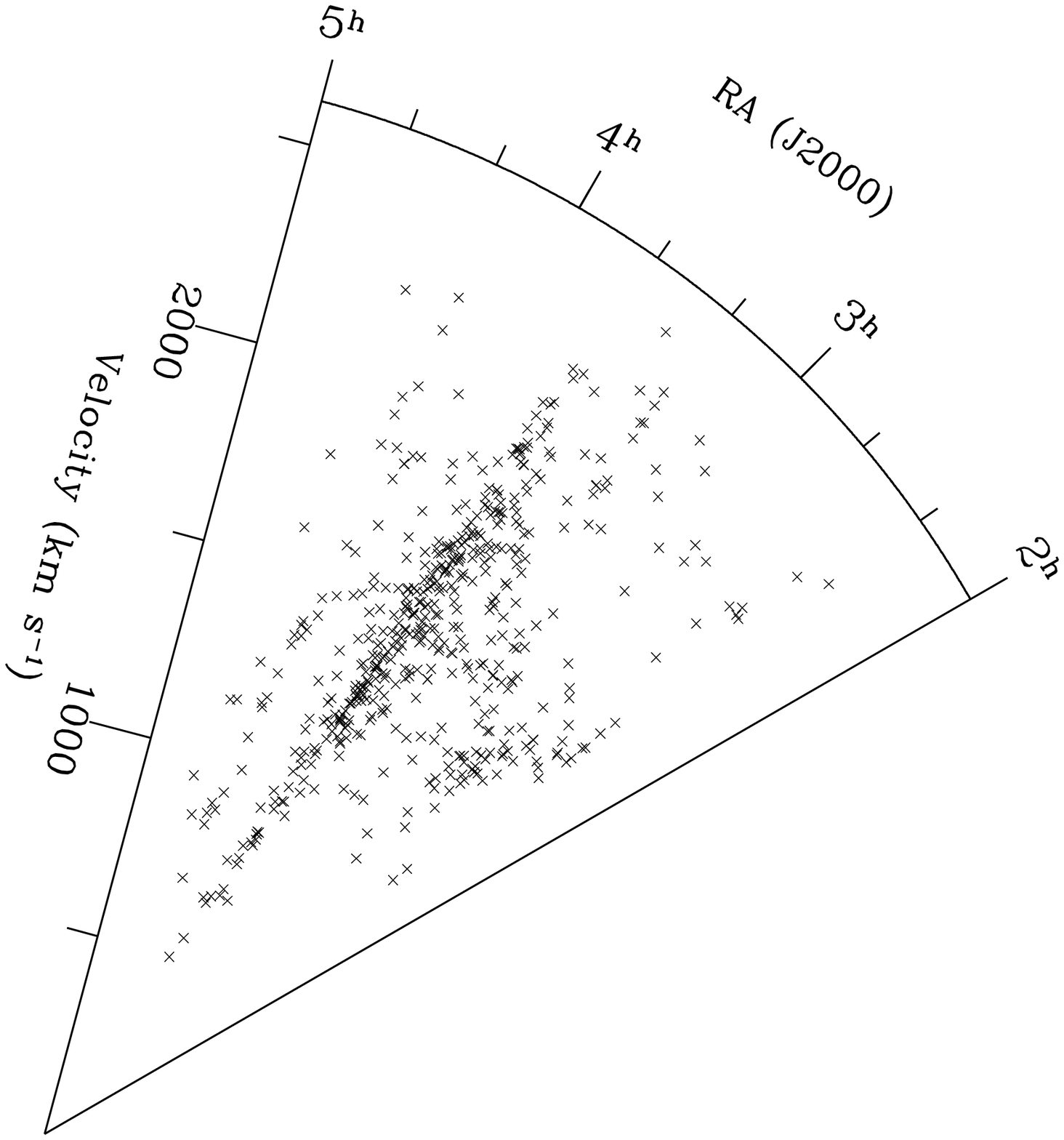}
    }
}
  \end{center}
\caption {Distribution of recession velocities of the 513 galaxies 
from both 6dFGS DR2 and NED in the Eridanus region.}
\label{coneplot}
\end{figure}

\begin{figure}
\begin{center}

    \resizebox{20pc}{!}{ \rotatebox{-90}{
    \includegraphics{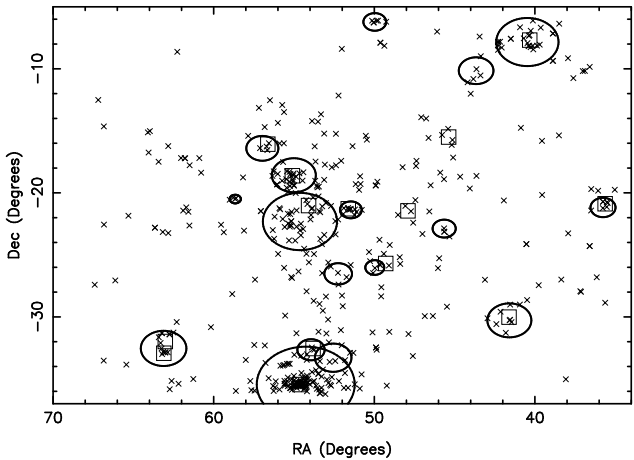} } }
    \end{center}
\caption {Spatial distribution of the 513 6dFGS DR2 and NED galaxies with 
$500 < v < 2500$ km s$^{-1}$.  The squares indicate the positions of
previously optically catalogued groups and the Fornax cluster (RA
$\sim 55$; Dec $\sim -35$) in this region with known velocities.  The
ellipses indicate the maximum radial extent of the 17 groups found with
the friends-of-friends algorithm }
\label{region}
\end{figure}

The 6dFGS is not yet complete.  
We therefore supplemented the 6dFGS data with sources from the
NASA/IPAC Extragalactic Database (NED) with known recession velocities
between 500 and $2500$ km s$^{-1}$.  The primary sources of these
velocities are: The Fornax compact object survey \citep{mieske04},
SDSS \citep{abazajian03}, 2dFGRS \citep{colless01} and the SSRS
\citep{dacosta98}. This added an additional 378 unique galaxies,
detailed in Appendix~\ref{ned_galaxies}.  Of these 378 galaxies, 266
either do not have 2MASS magnitudes or are fainter than the 2MASS
limit of $m_K\sim13.1$.  The total number of galaxies in the region
from NED and 6dFGS is 513.  Figure~\ref{coneplot} illustrates the
velocity distribution of these 513 galaxies, while Figure~\ref{region}
illustrates their spatial distribution.

\begin{figure}
\begin{center}

    \resizebox{20pc}{!}{
     \rotatebox{-90}{
	\includegraphics{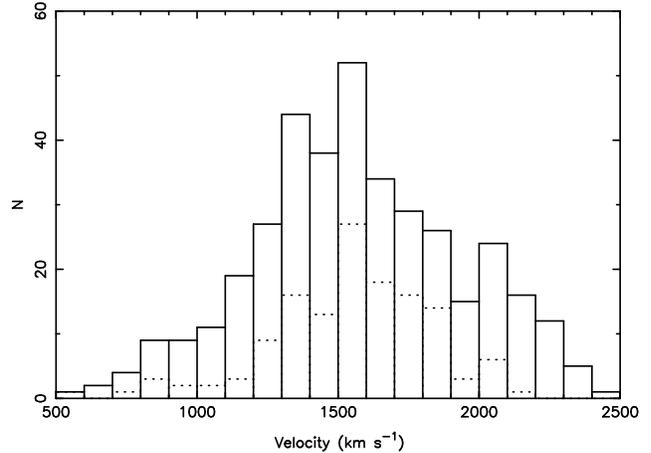}
    }
}
  \end{center}
\caption {Histogram of recession velocities of the 135 6dFGS DR2 
galaxies (dotted line) and 378 NED galaxies (solid line) in the
Eridanus region.}
\label{comp_vels}
\end{figure}

NED suffers from poorly contrained selection biases, it is therefore
important to illustrate that the distributions of recession velocities
of galaxies in the two samples are consistent with being drawn from
the same parent population.  Figure~\ref{comp_vels} compares the
recession velocities of both samples.  A Kolmogorov-Smirnov test finds
that the recession velocities from the 6dFGS and NED are consistent
with being drawn from the same parent population at the 96 per cent
level.

We also tested the effects of completeness by comparing our sample
with galaxies selected in the same region of sky and velocity from the
Hyper-Lyon-Meudon Extragalactic DAtabase (HyperLEDA;
\citealt{paturel97}).  HyperLEDA is a freely available database of 
observed galaxy quantities, including positions, velocities and
$B$-band magnitudes.  HyperLEDA has been shown to be photometrically
complete to at least $B\sim14$ mag (e.g. \citealt{giuricin00};
equivalent to $K\sim11$ mag).  We find all galaxies in HyperLEDA in
this region with $B\leq14$ mag to have velocities.  Therefore,
HyperLEDA is complete to $B\leq14$ mag in this region. We find our
sample to match that in HyperLEDA to this limit.  Therefore, our
sample is complete to at least $K\sim11$ mag.  We test the effects of
including galaxies fainter than this limit in our analyses in the
relevant sections.


We applied the limiting magnitude of the 2MASS data (i.e. $m_K<13.1$;
\citealt{jarrett00}), to the data for luminosity, morphology and colour 
analyses.  This created an apparent-magnitude limited sample of 201
galaxies.  Of these 201 galaxies, 89 are from the 6dFGS whilst 112 are
from NED.

\subsection{HyperLEDA}
Owing to the size of this region it was unfeasible to obtain new
photometric data so we used HyperLEDA to obtain total $B$-band
magnitudes and morphological T-types for the apparent
magnitude-limited sample.  T-types are numerical codes chosen to
correspond to morphological galaxy type.  The correspondance with
Hubble type is given in more detail in \cite{paturel97}.  In summary,
T-types of $-5\leq$ T-type $\leq0$ correspond to E to S0a galaxies
whilst $0<$ T-type $\leq10$ correspond to Sa to Irr galaxy types.
This resulted in 199 galaxies with measured $B$-band magnitudes and
193 with both $B$-band magnitudes and T-types.

\section{Analysis}
\label{section:analysis}

To avoid the effects of peculiar velocities which are significant at
recession velocities less than $2000$ km s$^{-1}$ \citep{marinoni98}
we use the distance modulus (DM; $m-M=31.6$) determined from the
globular cluster luminosity function of NGC 1407 \citep{forbes05}.
Using this DM, or that calculated from the surface brightness
fluctuation method of \cite{tonry01} which gives $m-M=31.84$ for the
Eridanus region (corrected following the work of \citealt{jensen03}),
has no effect on the general results presented below.  Absolute
magnitudes are calculated as $M=m-DM-A$. Galactic Extinction ($A$) is
calculated for the $K$- and $B$- bands using the extinction maps of
\cite{schlegel98} and is of the order $A_K\sim0.01$ mag and
$A_B\sim0.09$ mag.
Throughout this paper we assume $H_0=70$ km s$^{-1}$ Mpc$^{-1}$.



\subsection{Defining Structures}

In order to study the dynamics of the region it is important to
determine which galaxies are associated with each structure.  We used
the `friends-of-friends' (henceforth FOF; \citealt{huchra82})
percolation algorithm.  This method finds group structures in galaxy
data based on positional and velocity information and does not rely on
any {\it a priori} assumption about the geometrical shape of groups.
As we are examining a small range in recession velocities we do not
adopt the method used by \cite{huchra82} to compensate for the
sampling of the galaxy luminosity function as a function of the
distance of the group.



Owing to the similarity in sampling between the 2dFGRS and 6dFGS at
these recession velocities we follow the prescriptions of the 2dFGRS
Percolation-Inferred Galaxy Group (2PIGG;
\citealt{eke04}) catalogue to determine the most appropriate value of 
limiting density contrast, $\delta \rho/\rho$,  
\begin{equation}
\frac{\delta \rho}{\rho}=\frac{3}{4} \pi D_0^3\left[\int_{-\infty}^{M_{lim}}\phi(M)dM\right]^{-1}-1.
\end{equation}
The number density contour surrounding each group then represents a
fixed number density enhancement relative to the mean number density.
We assume the differential galaxy luminosity function defined by
\cite{kochanek01}, which \cite{ramella04} determine to be a good
approximation for the $K$-band groups luminosity function
($M_\star=-22.6$, $\alpha=-1.09$ and $\phi_\star=0.004$ for $H_0=70$
km s$^{-1}$ Mpc$^{-1}$).  The 2PIGG limiting density contrast $\delta
\rho/\rho=150$ then gives $D_0=0.29$ Mpc.  We also follow 2PIGG to
calculate our velocity limit, $V_0$.  The peculiar motion of galaxies
moving in a gravitational potential lengthens the group along the
line-of-sight in velocity space -- giving the `Finger of God'
effect.  If we assume that the projected spatial ($D_0$) and the
line-of-sight dimensions of a group in velocity space ($V_0$) are in
proportion, \cite{eke04} show that a ratio of 12 for $V_0$
relative to $D_0$
is the most appropriate for a linking volume, giving $V_0=347$ km
s$^{-1}$ here.


The FOF algorithm was run over the whole sample of 513 galaxies.  
We remove all groups with $N\leq3$ galaxies as these have been shown
by many surveys (e.g. \citealt{ramella95,nolthenius97,diaferio99}) to
be significantly more likely to be false positives found by the FOF
algorithm.

Following these methods we find 17 groups in the field, with an
average of 17.5 members per group.  These are illustrated in
Figure~\ref{region}.  We find 12 of the 14 groups previously defined
in this region.  Of the 2 we do not detect (USGC S108 and USGC S116)
both have only 4 members and have only been detected in the Updated
Zwicky Catalogue -- Southern Sky Redshift Survey 2 group catalogue
(UZC-SSRS2; \citealt{ramella02}).  The galaxies in these groups are
spatially extended suggesting that the UZC-SSRS2 group finder was
tuned to find looser structures than other group findesr used.

\begin{figure}
\begin{center}

    \resizebox{20pc}{!}{
     \rotatebox{-90}{
	\includegraphics{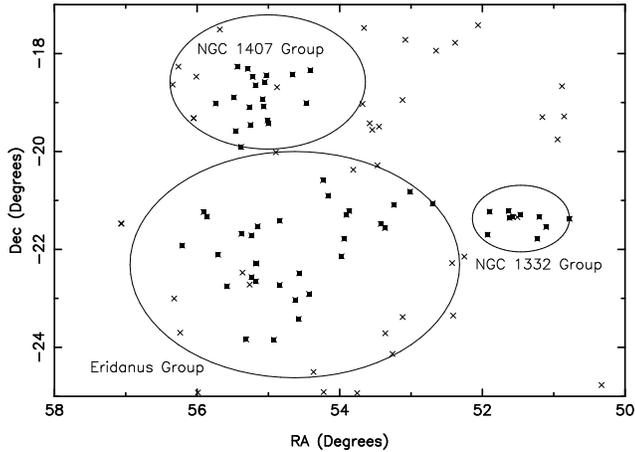}
    }
}
  \end{center}
\caption {The ellipses indicate the maximum radial extent of the 
groups found with the FOF algorithm (given in
Table~\ref{virial_table}).  The squares indicate the galaxies defined
as group members by FOF whilst crosses indicate non-group members.}
\label{fof}
\end{figure}

Those groups specifically associated with the Eridanus
region are illustrated in Figure~\ref{fof}.  This shows that the
algorithm finds distinct Eridanus, NGC 1407 and NGC 1332 groups.  The
galaxies in each group are detailed in Table~\ref{group_gals}.

\begin{table*}
\begin{center}
\caption{Details of the galaxies in each group.  $R_{gc}$ is the 
group-centric radius of each galaxy from the centre calculated by the
FOF algorithm.  Dashes indicate where information is not available.}
\label{group_gals}
\begin{tabular}{lccccccc}
\hline
Galaxy Name&6dFGS ID&RA (J2000)&Dec (J2000)&$v$ (km s$^{-1}$)&$m_K$ (mag)&$R_{gc}$ (Mpc)&T-type\\
\hline
NGC 1407 Group&&&&&&&\\
\hline
APMBGC 548-122-018          &  6dF J0341498-193453 & 3:41:49.82 & -19:34:52.5&  1914&    -      & 0.399 &   -    \\
ESO 548- G 076                &  6dF J0341318-195419 & 3:41:31.81 & -19:54:18.5&  1545&  11.877 & 0.501 & -1.3 \\
ESO 548- G 079                &  6dF J0341561-185343 & 3:41:56.08 & -18:53:42.6&  2031&  11.110 & 0.208 & -1.2 \\
ESO 548- G 073                &  6dF J0341044-190540 & 3:41: 4.41 & -19: 5:40.0&   987&  12.907 & 0.209 &  3.3 \\
2MASX J03401592-1904544       &  6dF J0340159-190454 & 3:40:15.93 & -19: 4:54.4&  1614&  12.262 & 0.183 & -3.5 \\
2MASX J03404323-1838431       &  6dF J0340432-183843 & 3:40:43.23 & -18:38:43.1&  1374&  12.364 & 0.067 & -1.6 \\
ESO 548- G 064                &  6dF J0340001-192535 & 3:40: 0.08 & -19:25:34.7&  1874&  11.032 & 0.307 & -2.7 \\
APMBGC 548-110-078          &  6dF J0340527-182841 & 3:40:52.73 & -18:28:40.8&  1680&    -      & 0.087 &    -  \\
NGC 1383                      &  6dF J0337392-182022 & 3:37:39.24 & -18:20:22.1&  2009&   9.506 & 0.234 & -1.9 \\
IC 0345                       &  6dF J0341091-181851 & 3:41: 9.13 & -18:18:50.9&  1245&  11.218 & 0.142 & -2.0 \\
NGC 1390                      &                      & 3:37:52.17 & -19: 0:30.1&  1207&  11.704 & 0.250 &  1.2 \\
NGC 1393                      &                      & 3:38:38.58 & -18:25:40.7&  2127&   9.312 & 0.138 & -1.7 \\
ESO 548- G 065                &                      & 3:40: 2.64 & -19:22: 0.7&  1221&  13.206 & 0.285 &  1.3 \\
IC 0343                       &                      & 3:40: 7.14 & -18:26:36.5&  1841&  10.647 & 0.052 & -0.8 \\
NGC 1407                      &                      & 3:40:11.84 & -18:34:48.5&  1779&   6.855 & 0.016 & -4.9 \\
ESO 548- G 068                &                      & 3:40:19.17 & -18:55:53.4&  1693&  10.427 & 0.129 & -2.6 \\
ESO 548- G 072                &                      & 3:41: 0.25 & -19:27:19.4&  2034&    -    & 0.330 &  5.0 \\
IC 0346                       &                      & 3:41:44.67 & -18:16: 1.2&  2013&  10.036 & 0.195 & -0.5 \\
ESO 549- G 002                &                      & 3:42:57.39 & -19: 1:14.9&  1111&    -    & 0.311 &  9.5 \\
\hline
Eridanus Group&&&&&&&\\
\hline
ESO 482- G 017                &  6dF J0337433-225430&  3:37:43.33  &-22:54:29.5 & 1515 & 12.748  &0.229  & 0.8   \\
LSBG F482-034                 &  6dF J0338166-222911&  3:38:16.55  &-22:29:11.4 & 1359 &   -     &0.067  &  -    \\
2MASX J03355395-2208228       &  6dF J0335540-220823&  3:35:53.95  &-22: 8:23.0 & 1374 & 12.737  &0.248  & 0.0   \\
2MASX J03354520-2146578       &  6dF J0335453-214659&  3:35:45.27 & -21:46:59.2&  1638&  13.678 & 0.318 &  0.0   \\
ESO 548- G 036                &  6dF J0333277-213353&  3:33:27.69 & -21:33:52.9&  1520&  10.541 & 0.536 &  0.1   \\
ESO 548- G 034                &  6dF J0332576-210522&  3:32:57.63 & -21: 5:21.9&  1707&  12.052 & 0.676 &  5.0   \\
NGC 1353                      &  6dF J0332030-204908&  3:32: 2.98 & -20:49: 8.2&  1587&   8.204 & 0.804 &  3.1   \\
2MASX J03365674-2035231       &  6dF J0336568-203523&  3:36:56.75 & -20:35:23.0&  1689&  12.422 & 0.645 &  0.0   \\
NGC 1377                      &  6dF J0336391-205407&  3:36:39.07 & -20:54: 7.2&  1809&   9.892 & 0.543 & -2.2   \\
ESO 548- G 069                &  6dF J0340362-213132&  3:40:36.17 & -21:31:32.4&  1647&    -    & 0.344 & 10.0   \\
NGC 1414                      &  6dF J0340571-214250&  3:40:57.14 & -21:42:49.9&  1752&    -    & 0.311 &  4.0   \\
APMUKS(BJ) B034114.27-212912  &  6dF J0343265-211944&  3:43:26.46 & -21:19:44.2&  1711&    -    & 0.574 &   -    \\
NGC 1422                      &  6dF J0341311-214054&  3:41:31.07 & -21:40:53.5&  1680&  10.973 & 0.357 &  2.3   \\
NGC 1415                      &  6dF J0340569-223352&  3:40:56.86 & -22:33:52.1&  1659&   8.388 & 0.239 &  0.5   \\
ESO 482- G 031                &  6dF J0340415-223904&  3:40:41.54 & -22:39: 4.1&  1803&  12.806 & 0.233 & -1.9   \\
ESO 548- G 029                &                     &  3:30:47.17 & -21: 3:29.6&  1215&  11.214 & 0.841 &  3.4   \\
IC 1953                       &                     &  3:33:41.87 & -21:28:43.1&  1867&  10.100 & 0.535 &  6.7   \\
ESO 548- G 049                &                     &  3:35:28.27 & -21:13: 2.2&  1510&    -    & 0.487 &  5.6   \\
IC 1962                       &                     &  3:35:37.38 & -21:17:36.8&  1806&    -    & 0.457 &  7.5   \\
ESO 482- G 018                &                     &  3:38:17.64 & -23:25: 9.0&  1687&  11.765 & 0.403 &  0.3   \\
NGC 1395                      &                     &  3:38:29.72 & -23: 1:38.7&  1717&   7.024 & 0.260 & -5.0   \\
MCG -04-09-043                &                     &  3:39:21.57 & -21:24:54.6&  1588&    -    & 0.337 &  2.9   \\
NGC 1401                      &                     &  3:39:21.85 & -22:43:28.9&  1495&   9.453 & 0.168 & -2.1   \\
ESO 482- G 035                &                     &  3:41:14.65 & -23:50:19.9&  1890&  10.967 & 0.609 &  2.3   \\
NGC 1426                      &                     &  3:42:49.11 & -22: 6:30.1&  1443&   8.762 & 0.399 & -4.8   \\
ESO 549- G 006                &                     &  3:43:38.25 & -21:14:13.7&  1609&    -    & 0.609 &  9.7   \\
NGC 1439                      &                     &  3:44:49.95 & -21:55:14.0&  1670&   8.728 & 0.593 & -5.0   \\
APMUKS(BJ) B033830.70-222643  &                     &  3:40:41.35 & -22:17:10.5&  1737&    -    & 0.198 &   -      \\
ESO 482- G 027                &                     &  3:39:41.21  &-23:50:39.8 & 1626 &   -     &0.568  &10.0  \\
ESO 548- G 070                &                     &  3:40:40.99  &-22:17:13.4 & 1422 &   -     &0.197  & 7.0  \\
ESO 482- G 036                &                     &  3:42:18.80  &-22:45: 9.2 & 1567 &   -     &0.381  & 9.1   \\
\hline
NGC 1332 Group&&&&&&&\\
\hline
NGC 1315                      & 6dF J0323066-212231 & 3:23: 6.60 & -21:22:30.7&  1597&   9.944 & 0.248 & -1.0  \\
2MASX J03255262-2117204       & 6dF J0325526-211721 & 3:25:52.62 & -21:17:20.6&  1428&  11.767 & 0.029 &   -   \\
NGC 1331                      & 6dF J0326283-212120 & 3:26:28.34 & -21:21:20.3&  1242&  10.867 & 0.058 & -4.7  \\
NGC 1325                      & 6dF J0324256-213238 & 3:24:25.57 & -21:32:38.3&  1590&   8.831 & 0.143 &  4.2  \\
ESO 548- G 022                & 6dF J0327422-214159 & 3:27:42.16 & -21:41:58.6&  1295&    -    & 0.209 &  5.0  \\
ESO 548- G 021                & 6dF J0327356-211341 & 3:27:35.57 & -21:13:41.4&  1745&    -    & 0.168 &  7.7  \\
NGC 1325A                     &                     & 3:24:48.50 & -21:20:11.5&  1333&  14.035 & 0.094 &  6.6  \\
ESO 548- G 011                &                     & 3:24:55.31 & -21:47: 0.6&  1453&    -    & 0.173 &  8.4  \\
NGC 1332                      &                     & 3:26:17.25 & -21:20: 7.2&  1524&   7.122 & 0.043 & -3.1  \\
2MASX J03263135-2113003       &                     & 3:26:31.34 & -21:13: 0.5&  1548&  11.263 & 0.084 & -2.5  \\
\hline
\end{tabular} 
\end{center}
\end{table*}

\section{Dynamics}
\label{sect_dyn}
For each group the luminosity-weighted centroid and mean recession
velocity were calculated.  These and the dynamical parameters
calculated using the group members defined by the FOF algorithm are
summarised in Table~\ref{virial_table}.  

\begin{figure*}
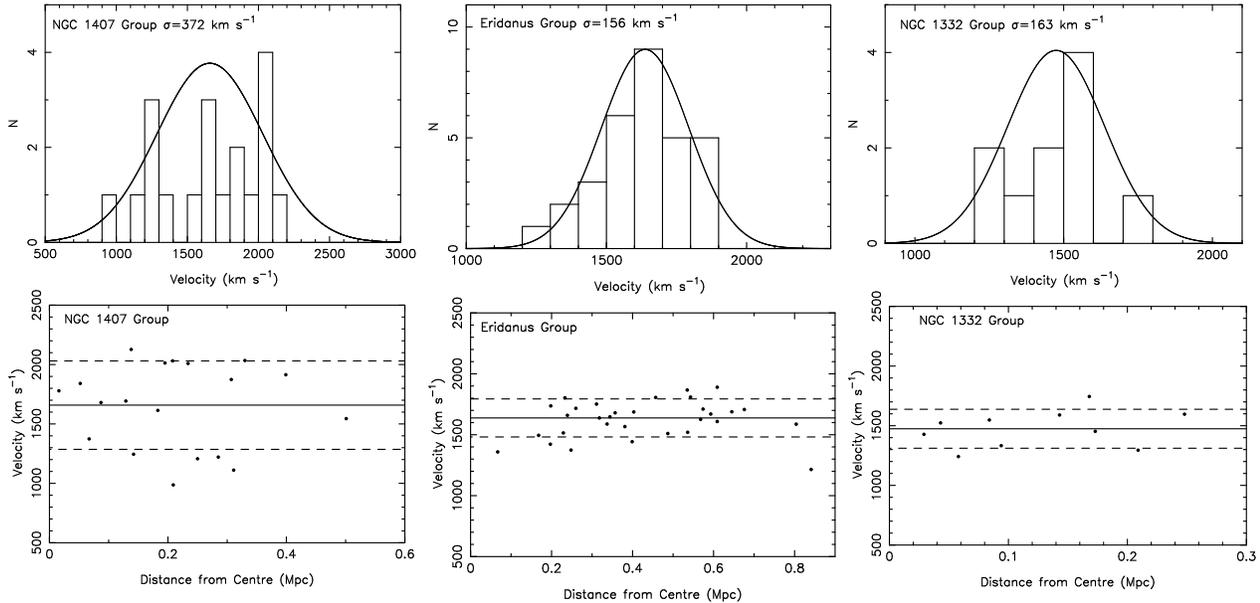

\begin{center}

    \resizebox{13pc}{!}{
     \rotatebox{-90}{
	\includegraphics{velocity_histo_1407.ps}
	\includegraphics{vd_1407.ps}
}
}
    \resizebox{13pc}{!}{
     \rotatebox{-90}{
	\includegraphics{velocity_histo_eridanus.ps}
	\includegraphics{vd_erid.ps}
}
}
    \resizebox{13pc}{!}{
     \rotatebox{-90}{
	\includegraphics{velocity_histo_1332.ps}
	\includegraphics{vd_1332.ps}
    }
}

  \end{center}
\caption{Velocity distributions of the three groups as defined by 
the FOF algorithm.  The upper panel shows the velocity histograms with
the line representing a Gaussian with $\sigma_v=$ velocity dispersion of
each group.  The lower panel shows the velocity-distance relationships
for each group.  The solid line represents the mean velocity whilst
the dashed lines indicate the velocity dispersions.}
\label{vel_histo}
\end{figure*}

The velocity dispersion, $\sigma_v$, was calculated using the gapper
algorithm.  This is insensitive to outliers, giving a robust estimate
of $\sigma_v$ for small groups \citep{beers90}.

\begin{equation}
\sigma_v=c \sqrt \frac{\pi}{[n(n-1)]} \sum_{i=1}^{n-1} w_i g_i,
\end{equation}
where $w_i=i(n-i)$ and $g_i=z(i+1)-z(i)$.  The corresponding errors
are estimated using the jackknife algorithm.  The velocity
distribution of each group with the calculated velocity dispersion
overlaid are illustrated in Figure~\ref{vel_histo}.

If these groups are not yet relaxed then there may be evidence of that
in their velocity distributions.  We examine the higher moments of the
velocity distributions using the kurtosis and skew.  Kurtosis
indicates a difference in the length of tails of the distribution
compared to a Gaussian.  Zero values indicate a Gaussian distribution
whilst positive kurtosis indicates longer tails than a Gaussian
distribution.  The standard deviation on this quantity is given by
$\sqrt(24/N)$.  Skewness indicates asymmetry, where zero again
indicates Gaussianity and positive skewness implies that the
distribution is depleted from values lower than the mean location.
The standard deviation on this quantity is given by $\sqrt(6/N)$.
Significant deviations from a Gaussian distribution are defined by
values greater than the standard deviations on these quantities.  All
of our groups show large deviations such that none of the groups show
significant signs of skewness or kurtosis.  We conclude that their
velocity distributions are consistent with their being virialized
structures.

The crossing time is calculated as a function of the Hubble time
($H_0^{-1}$) following \cite{huchra82}, as:
\begin{equation}
t_c=\frac{3~r_H}{5^{3/2}\sigma_v},
\end{equation}
where the harmonic radius, $r_H$, is independent of the velocity
dispersion and is given below.  
\begin{equation}
r_H=\pi~D~sin~\left[\frac{n(n-1)}{4 \sum_i \sum_{j>i} \theta_{ij}^{-1}}\right],
\end{equation}
where $D$ is the distance to the group from the distance modulus of
\cite{forbes05} 
i.e. $D=20.89$ Mpc, $n$ is the number of members of each group, and
$\theta_{ij}^{-1}$ is the angular separation of group members.  We
calculated the error on $r_H$ using the jackknife method and then used
standard error propagation analysis to calculate the rms error on the
crossing time.  \cite{nolthenius87} indicate that a crossing time
$>0.09$ H$_0^{-1}$ suggests that a group has not yet had time to
virialize.

The virial mass $M_V$ was calculated using the virial mass 
estimator of \cite{heisler85}.
\begin{equation}
M_V=\frac{3~\pi~N}{2~G}\frac{\sum_i V_i^2} {\sum_{i<j}1/R_{gc,ij}},
\label{virial_mass}
\end{equation}
where $V_i$ is the observed radial component of the velocity of the
galaxy $i$ with respect to the systemic group velocity and $R_{gc,i}$
is its projected separation from the group centre.  We estimate the
rms error of the mass using the jacknife method (e.g. \citealt{biviano93}).

The radius corresponding to an overdensity of 500 times the critical
density -- $r_{500}$, is calculated as a function of the velocity
dispersion following \cite{osmond04} as,
\begin{equation}
r_{500}=\frac{0.096\sigma_v}{H_0}.
\end{equation}
The rms error on the $r_{500}$ values were estimated by standard error
propagation.  In Table~\ref{virial_table} we also give the X-ray
luminosities for these groups.  For the NGC 1407 and NGC 1332 groups
these were calculated by \cite{osmond04} using their $r_{500}$ values.
The X-ray luminosity for the Eridanus group is calculated by
\cite{omar05a} for NGC 1395 in Eridanus.

The mass-to-light ratios were calculated by dividing the virial mass
by the sum of the luminosities of the member galaxies for both the
$K$- and $B$-bands.  The errors were estimated using standard error
propagation.

The spiral fraction, $f_{sp}$, was calculated as the fraction of
galaxies in each group with $m_K<13.1$ mag and T-type $>0.0$.  The
errors quoted in Table~\ref{virial_table} are the poisson errors on
this value.

\begin{table}
\begin{center}
\caption{Derived group properties.  The number of members in brackets are 
those with $m_K<13.1$ mag.  See text for details on parameters.}
\label{virial_table}
\begin{tabular}{|c|ccc|}
\hline
&NGC 1407&Eridanus&NGC 1332\\ \hline 
Centroid (J2000)&03:40:02,&03:38:32,&03:25:50,\\
&-18:35:03&-22:18:51&-21:22:8\\ 
Members&19 (14)&31 (18)&10 (6)\\ 
$v$ (km s$^{-1}$)&1658$\pm26$&1638$\pm5$&1474$\pm18$\\
Max. radial extent (Mpc)&0.5&0.8&0.3\\ 
$\sigma_v$ (km s$^{-1}$)&372$\pm48$&156$\pm23$&163$\pm35$\\
Skewness&-0.38$\pm$0.56&-0.53$\pm$0.44&0.05$\pm$0.77\\
Kurtosis&-1.36$\pm$1.12&-0.14$\pm$0.88&-1.28$\pm$1.55\\ 
$t_c$(H$_0^{-1}$)&0.02$\pm0.005$&0.03$\pm0.6$&0.02$\pm0.1$\\
$M_V~(10^{13}$M$_{\odot})$&5.3$\pm1.5$&0.9$\pm1.5$&0.6$\pm0.4$\\ 
$r_{500}$ (Mpc)&0.51$\pm0.07$&0.21$\pm0.03$&0.22$\pm0.05$\\ 
Log $L_X (r_{500})$ (erg s$^{-1})$&41.92$\pm$0.02&40.83&40.93$\pm$0.02\\ 
$M_V/L_K$&230$\pm63$&27$\pm44$&35$\pm21$\\
$f_{sp}$&0.14$\pm0.04$&0.56$\pm0.13$&0.20$\pm0.09$\\
\hline
\end{tabular} 
\end{center}
\end{table}

\begin{figure*}
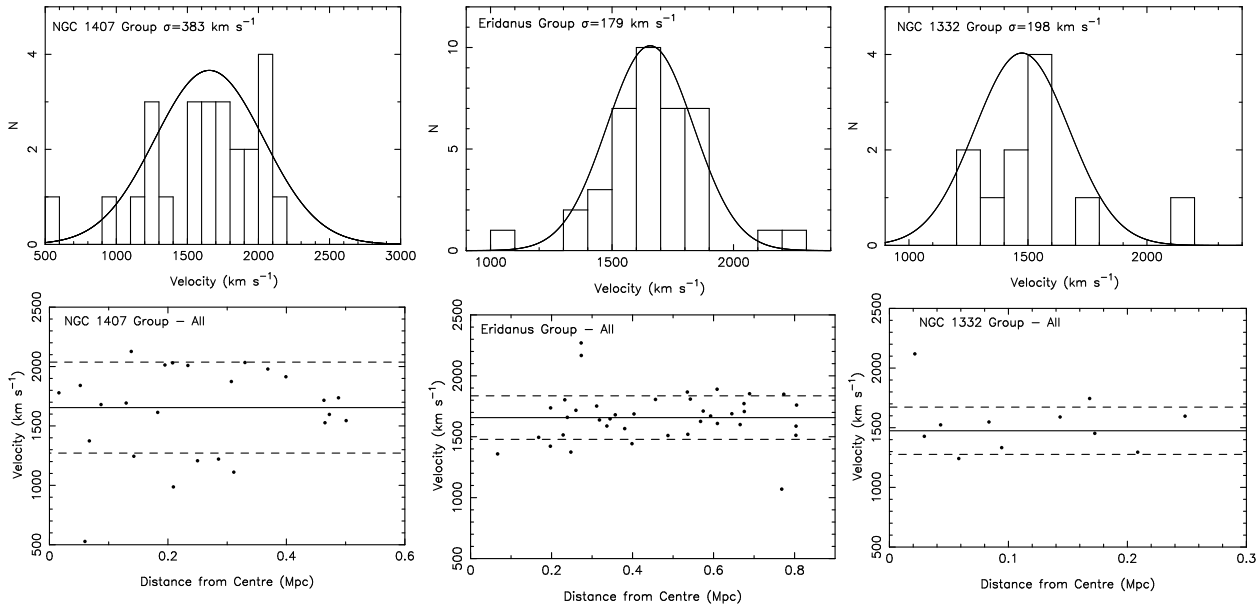

\begin{center}

    \resizebox{13pc}{!}{
     \rotatebox{-90}{
	\includegraphics{histo_1407_all.ps}	
	\includegraphics{vd_1407_all.ps}
}
}
    \resizebox{13pc}{!}{
     \rotatebox{-90}{
	\includegraphics{histo_erid_all.ps}	
	\includegraphics{vd_erid_all.ps}
}
}
    \resizebox{13pc}{!}{
     \rotatebox{-90}{
	\includegraphics{histo_1332_all.ps}
	\includegraphics{vd_1332_all.ps}
    }
}

  \end{center}
\caption{Velocity distributions of the three groups consisting 
of all galaxies within the maximum radial extent of the groups defined
by FOF.  The upper panel shows the velocity histograms with
the line representing a Gaussian with $\sigma_v=$ velocity dispersion of
each group.  The lower panel shows the velocity-distance relationships
for each group.  The solid line represents the mean velocity whilst
the dashed lines indicate the velocity dispersions.}
\label{vel_histo_all}
\end{figure*}

In order to determine how the FOF definition of membership affects the
dynamical properties of these groups, we also examined the properties
of groups consisting of {\it all} galaxies within the maximum radial
extent of the groups defined by FOF.  Their velocity-distance
relationships are illustrated in Figure~\ref{vel_histo_all}, and the
associated parameters given in Table~\ref{virial_table_all}.

It is clear from Table~\ref{virial_table_all} that the derived mean
recession velocities, velocity dispersions and radii are, within the
errors, consistent with those determined using the FOF membership and
are clearly robust to the definition of group membership.  However,
when all galaxies are considered, the velocity dispersions are larger
than the FOF measurements, hence the virial masses are larger


\begin{table}
\begin{center}
\caption{Derived properties for groups consisting of all galaxies 
within the radial extent of the groups defined by FOF.}
\label{virial_table_all}
\begin{tabular}{|c|ccc|}
\hline
&NGC 1407&Eridanus&NGC 1332\\ \hline 
Members&25&39&11\\ 
$v$ (km s$^{-1}$)&1652$\pm19$&1657$\pm4$&1475$\pm22$\\
$\sigma_v$ (km s$^{-1}$)&384$\pm63$&179$\pm33$&198$\pm80$\\
$M_V~(10^{13}$M$_{\odot})$&7.9$\pm2.8$&2.1$\pm2.7$&1.4$\pm0.9$\\ 
$r_{500}$ (Mpc)&0.53$\pm0.09$&0.24$\pm0.05$&0.27$\pm0.1$\\ 
\hline
\end{tabular} 
\end{center}
\end{table}

\subsection{Evaluating the Effects of Incompleteness}

We tested for the effects of incompleteness on our analysis by
reducing the sample to those 121 galaxies with $m_K\leq11$ mag, at
which point we know that our sample is complete.  We then re-ran our
group-finder and re-calculated the dynamical parameters for each
group.  We iterated this process increasing the limiting magnitude,
$m_{K,lim}$, by $\sim0.5$ mag each time.

At each iteration, three individual groups are found in the supergroup
region.  Therefore, the structure we are examining is robust to the
effects of incompleteness.

\begin{figure}
\begin{center}

    \resizebox{20pc}{!}{
     \rotatebox{-90}{
	\includegraphics{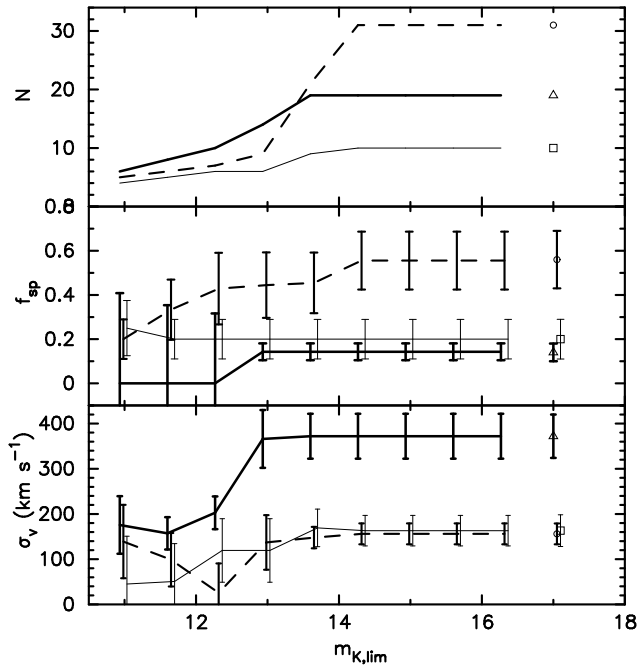}	
}}
  \end{center}
\caption{The relationship of group members, $N$, spiral fraction, 
$f_{sp}$, and velocity dispersion, $\sigma_v$, with increasing
limiting apparent magnitude, $m_{K,lim}$, for the NGC 1332 (thin
lines), Eridanus (dashed lines) and NGC 1407 (thick lines) groups.
The x-axis positions of the NGC 1332 and Eridanus points are shifted
for clarity.  The error bars indicate $1\sigma$ errors, as calculated
in Section~\ref{sect_dyn}.  The points and their associated errors
give the values found for each parameter using the whole dataset,
presented in Table~\ref{virial_table}: Squares represent the NGC 1332
group, circles represent Eridanus, and triangles represent the NGC
1407 group.}
\label{limit_check_all}
\end{figure}

In Figure~\ref{limit_check_all}, we illustrate how the dynamical
parameters of these three groups vary with increasing magnitude limit.
We study the number of galaxies in each group, the spiral fractions in
those groups ($\sim$ galaxy populations) and the velocity dispersions
of those groups ($\sim$ group mass).  It is worth noting that all
three parameters of all three groups reach a plateau at
$m_{K,lim}\sim13-14$ mag, the 2MASS apparent-magnitude limit.
Examining Figure~\ref{limit_check_all} it is clear that increasing the
number of galaxies in the sample increases the number of galaxies in
each group.  The spiral fraction in each group also rises, although
the spiral fractions calculated for a limiting magnitude
$m_{K,lim}\sim11$ mag are, within the errors, consistent with those
calculated for the whole sample.  The velocity dispersion also rises
with an increasing number of galaxies in the sample.  For the NGC 1332
and Eridanus groups the velocity dispersions calculated for a limiting
magnitude $m_{K,lim}\sim11$ mag are, within the errors, consistent
with those calculated for the whole sample.  However, for the NGC 1407
group the velocity dispersion of the whole sample is significantly
larger than that calculated for the sample limited to
$m_{K,lim}\sim11$ mag.  It is known that velocity dispersions
calculated for undersampled groups are lower than they are in reality.
It is therefore difficult to determine whether the significant change
in the velocity dispersion of the NGC 1407 group with increasing
magnitude limit is due to undersampling in the sample limited to
$m_{K}\leq11$ or incompleteness in the whole sample.  We test the
effects of incompleteness by randomly removing a percentage of the
dataset with magnitudes $>11$ mag.  We repeat the random removal 1000
times, and calculate the mean difference $\langle
\sigma_{v,all}-\sigma_{v,lim} \rangle$ and the error on that mean.  For
the NGC 1407 group, removing 10 per cent of the faint-end of the
sample results in $\langle \sigma_{v,all}-\sigma_{v,lim}
\rangle=4.2\pm0.5$ (km s$^{-1}$), removing 20 per cent $\langle
\sigma_{v,all}-\sigma_{v,lim} \rangle=9.2\pm1.2$ (km s$^{-1}$), and
removing 40 per cent $\langle \sigma_{v,all}-\sigma_{v,lim}
\rangle=51\pm3$ (km s$^{-1}$).  These values are all significantly lower 
than the observed difference and therefore the increase in $\sigma_v$
with increasing limiting magnitude cannot be due to incompleteness in
the sample.  Rather it is more likely to be a result of better
sampling of this group due to larger number statistics.  We therefore
conclude that the results we present are robust to the effects of
incompleteness.

\subsection{Individual Group Properties}
\subsubsection{NGC 1407 Group}

The NGC 1407 group centroid defined by our FOF analysis is 121 km
s$^{-1}$ and 16 kpc away from the brightest galaxy in the group, the
large elliptical NGC 1407.  This difference is within the error of the
mean recession velocity and the position of NGC 1407 itself.  Thus,
the brightest group galaxy, NGC 1407, lies at the spatial and
kinematic centre of the group.
The numbers of galaxies and mean velocity found are close to that
determined by \cite{osmond04}, who find $N=20;~v=1489\pm59$ km
s$^{-1}$. The FOF algorithm does not find NGC 1400 (at a recession
velocity of $528$ km s$^{-1}$) to be a member of this group.  This is
in contrast to previous work
(e.g. \citealt{gould93,quintana94,perrett97,tonry01,forbes05}) that
suggests that this galaxy is actually at the same distance as NGC
1407.  Table~\ref{virial_table_all} shows that including NGC 1400 in
the group does not have a significant effect on the derived velocity
dispersion.


The velocity dispersion measured here is consistent with that measured
by \cite{osmond04}, i.e. $\sigma_v=319\pm52$ km s$^{-1}$.
\cite{miles04} observe a dip in the luminosity function of groups with
log $L_X<41.7$ erg s$^{-1}$ and conclude that such groups are the
present sites of rapid merging. The X-ray luminosity of the NGC 1407
group is higher than this, suggesting that it is unlikely to be
undergoing rapid dynamical evolution at this time.  The
high mass-to-light ratio, low kurtosis and skewness, low spiral
fraction, symmetric intra-group X-ray emission, bright central
elliptical galaxy and short crossing time all indicate that this
structure is virialized.

\subsubsection{Eridanus Group}

In contrast to the NGC 1407 group, the Eridanus group is not centred
on any one galaxy.  The brightest galaxy in the group, the large
elliptical NGC 1395 with its galaxy group-sized X-ray halo, is 300 kpc
and 79 km s$^{-1}$ away from the centre defined by the FOF code.  The
Eridanus group is made up of more galaxies than the NGC 1407 group
($N=31$) but is a much looser, irregular structure.  This is echoed in
its high spiral fraction and low velocity dispersion. 


\subsubsection{NGC 1332 Group}

The NGC 1332 group centroid is $50$ km s$^{-1}$ and 43 kpc from the
position of the brightest galaxy in the group, the large lenticular
NGC 1332 and its associated X-ray emission.  This is within the error
of the mean recession velocity and the position of NGC 1332.  Thus,
similar to the NGC 1407 group, the brightest group galaxy, NGC 1332,
lies at the spatial and kinematic centre of the group.
This group has fewer galaxies ($N=10$) than either NGC 1407 or
Eridanus.  \cite{osmond04} found the same number of galaxies
associated with this group at a similar recession velocity
($v=1489\pm59$ km s$^{-1}$).
The measured velocity dispersion is within the errors of that measured
by \cite{osmond04}, i.e. $\sigma_v=186\pm45$ km s$^{-1}$.  The measured
velocity dispersion is small, consistent with it being a low-mass
group.  
Its skewness and kurtosis are consistent with virialisation.  Its
spiral fraction is similar to that of the NGC 1407 group.  These
properties suggest that NGC 1332 group is a low-mass, compact,
virialized group with a galaxy population similar to that of a much
more massive group like NGC 1407.  However, its lack of intra-group
X-ray emission suggests that it is not as dynamically mature as the
NGC 1407 group.

\subsection{The Eridanus Region -- A Supergroup?}

A supergroup is a 
group of groups that will eventually merge to form a cluster.
In order to determine whether the three groups in the Eridanus region
form a supergroup it is important to establish whether the groups are
actually bound to one another.

\begin{figure*}
\begin{center}

    \resizebox{30pc}{!}{
     \rotatebox{-90}{
	\includegraphics{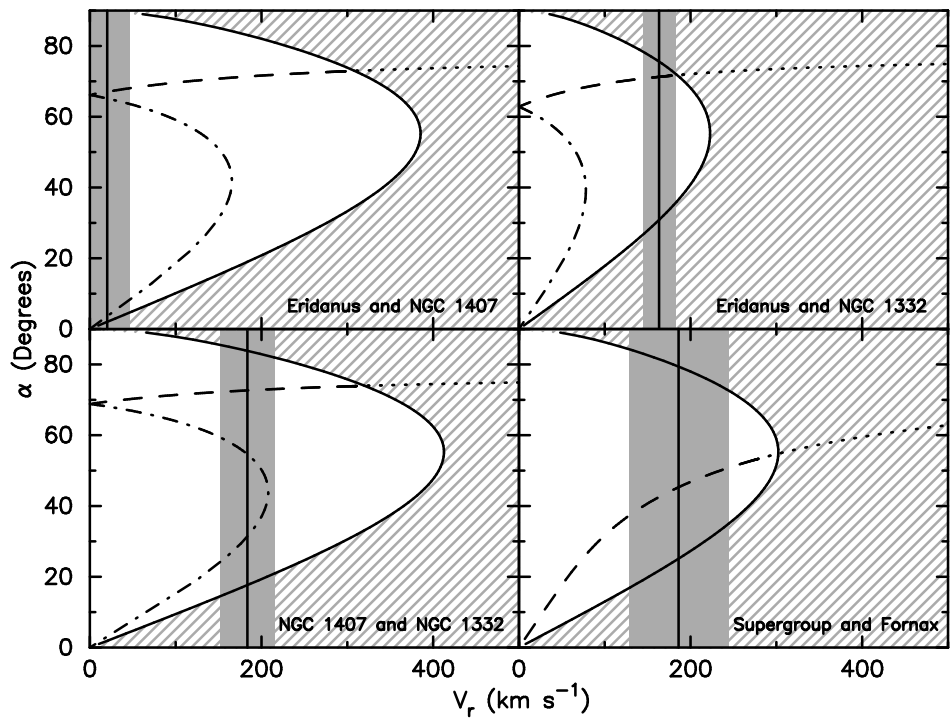}	
}
}
  \end{center}
\caption{Projection angle, $\alpha$, as a function of the relative velocity 
between two groups, $V_r$.  The hashed region marks the unbound
solutions, the dotted line marking the unbound orbit lies in this
region.  The dashed lines mark the bound-outgoing and the dot-dashed
lines mark the bound-incoming solutions, while the vertical straight
line and shaded region indicate the observed relative velocity between
the groups and their uncertainty.  The group pairs studied are stated
in each panel.  }
\label{vel_rel_all}
\end{figure*}

We use the Newtonian binding criterion that a two-body system is bound
if the potential energy of the bound system is equal to or greater
than the kinetic energy.  To assess the likelihood that the individual
groups are bound to one another we require
(e.g. \citealt{beers82,gregory84,cortese04}):  

\begin{equation}
V_r^2R_p\leq 2 G M \sin^2\alpha~\cos \alpha,
\label{bound_eq}
\end{equation}
where $M$ is the total mass of the system, $V_r$ is the relative
velocity between the groups, sin$~\alpha$ corrects $V_r$ for the
projection effects, $R_p$ is the projected distance between the
groups and cos$~\alpha$ corrects $R_p$ for projection.  We analyse
the three pairs of groups separately and give the variables used in
the analysis for each pair in Table~\ref{bound_table}.  The hashed
regions in Figure~\ref{vel_rel_all} illustrate the solutions to
Equation~\ref{bound_eq} for each pair of groups.

Considering the solid angles over which the systems are bound, given
the observed relative velocity, we calculate the probabilities that
each two-body system is bound following \cite{girardi05}:
\begin{equation}
P_{bound}=\int^{a2}_{a1}\cos \alpha ~d\alpha, 
\end{equation}
where the system is bound between angles $a1$ and $a2$ given the
observed relative velocity.  These probabilities are given for each
pair in Table~\ref{bound_table}.  Whilst the Eridanus and NGC 1407
pair is likely to be bound at the 96 per cent level, and the NGC 1407
and NGC 1332 pair at the 70 per cent level, the Eridanus and NGC 1332
pair are only likely to be bound at the 40 per cent level.  As the
conservative estimate that the three groups are bound is based on the
minimum probability that each of the three pairs are bound, the low
probability of the Eridanus and NGC 1332 pair being bound means that
this system is not bound.  However, individual groups are clearly
showing close relationships.

\begin{table*}
\begin{center}
\caption{Dynamical properties of the group pairs, used in the two-body 
binding analysis.  The parameters are described in the text.}
\label{bound_table}
\begin{tabular}{cccccccc}
\hline
Group Pair&$M$&$V_r$&$R_p$&$P_{bound}$&$P_{BO}$&$P_{BIa}$&$P_{BOa}$\\ 
&$(10^{13}M_{\odot})$&(km s$^{-1}$)&(Mpc)&(per cent)&(per cent)&(per cent)&(per cent)\\
\hline
Eridanus \& NGC 1407&6.2&20$\pm$26&1.6&96.5&1.8&31&68\\
Eridanus \& NGC 1332&1.5&163$\pm19$&1.2&46&100&-&-\\
NGC 1407 \& NGC 1332&5.9&184$\pm32$&1.4&70&1.3&44&55\\
Supergroup \& Fornax&13.7&186$\pm58$&5.9&56&100&-&-\\
\hline
\end{tabular} 
\end{center}
\end{table*}

We study this further by applying the two-body analysis described by
\cite{beers82} to each pair.  The two bodies are treated as point
masses moving on radial orbits.  They are assumed to start their
evolution at time $t=0$ with no separation and are moving apart or
coming together for the first time in their history.  For bound radial
orbits, the parametric solutions to the equations of motion are:

\begin{equation}
r=\frac{R_m}{2}(1-\cos \chi),
\end{equation}
\begin{equation}
t=\left(\frac{R_m^3}{8GM}\right)^{1/2}(\chi-\sin \chi),
\end{equation}
\begin{equation}
V=\left(\frac{2GM}{R_m}\right)^{1/2}\frac{\sin \chi}{(1-\cos \chi)},
\end{equation}
where $r$ is the components separation at time $t$, $R_m$ is the
separation of the components at maximum expansion, $\chi$ is the
developmental angle and $V$ is their real, relative velocity.  The
observables $V_r$ and $R_p$ are related to the real parameters by:

\begin{equation}
R_p=r~\cos \alpha,~V_r=V~ \sin \alpha.
\end{equation}

We close the equations by setting $t=13.7$ Gyrs, the age of the
Universe in a $\Lambda$CDM cosmology.  We can then solve the equations
above for $V_r$ and $\alpha$ using Equation 6 from \cite{gregory84}:

\begin{equation}
\tan \alpha=\frac{tV_r}{R_p}\frac{(\cos \chi -1)^2}{\sin \chi(\chi-\sin \chi)}.
\end{equation}

These solutions are illustrated by the dashed (bound-outgoing; $BO$)
and dot-dashed (bound-incoming; $BI$) lines in
Figure~\ref{vel_rel_all}.  For some pairs of groups there are two
solutions to the bound-incoming case within our observed relative
velocities, owing to the ambiguity in the projection angle, $\alpha$.
These are denoted $BI_a$ and $BI_b$.  We calculate the probabilities
$P_{BO}$, $P_{BIa} $ and $P_{BOa}$ again following \cite{girardi05} by
considering the range of solutions that are consistent with the
observed relative velocities.  Therefore,
\begin{equation}
P_{BO}=p_{BO}/\Sigma p=[\sin (a4)-\sin (a3)]/\Sigma p,
\end{equation}
\begin{equation}
P_{BIa}=p_{BIa}/\Sigma p=[\sin (a6)-\sin (a5)]/\Sigma p,
\end{equation}
\begin{equation}
P_{BIb}=p_{BIb}/\Sigma p=[\sin (a7)-\sin (a6)]/\Sigma p,
\end{equation}
where the angles $a3$ etc. are the angular limits of the solutions
and,
\begin{equation}
\Sigma p=p_{BO}+p_{BIa}+p_{BIb}.
\end{equation}

These probabilities are also given in Table~\ref{bound_table}.  We
find that the Eridanus and NGC 1407 pair is likely to be bound and
incoming (i.e. contracting) at the 98.2 per cent level.  The NGC 1407
and NGC 1332 group pair is also likely to be bound and incoming at the
98.7 per cent level.  The bound-incoming solution for the Eridanus and
NGC 1332 pair is unlikely given the observed relative velocities
between the two groups, the bound-outgoing solution is therefore the
most likely.

\cite{beers82}, \cite{gregory84} and \cite{cortese04} also provide the 
parametric solutions for the unbound case:

\begin{equation}
r=\frac{GM}{V_{\infty}^2}(\cosh \chi -1),
\end{equation}
\begin{equation}
t=\frac{GM}{V_{\infty}^3}(\sinh \chi-\chi),
\end{equation}
\begin{equation}
V=V_{\infty}\frac{\sinh \chi}{\cosh \chi-1},
\end{equation}
where $V_{\infty}$ is the asymptotic expansion velocity.  As for the
bound case, we can then solve the equations above for $V_r$ and
$\alpha$ using Equation 6 from \cite{gregory84}:
\begin{equation}
\tan \alpha=\frac{tV_r}{R_p}\frac{(\cosh \chi -1)^2}{\sinh \chi(\sinh \chi-\chi)}.
\end{equation}

The solutions to the unbound case are plotted in
Figure~\ref{vel_rel_all} as dotted lines.  The relative velocities we
observe are clearly inconsistent with the unbound orbits in all
cases.  

The calculated probabilities show that it is unlikely that the three
Eridanus groups form a gravitationally bound structure.  However, the
Eridanus and NGC 1407 groups and, at a lesser level, the NGC 1407 and
NGC 1332 groups are individually bound.  The two-body orbital analysis
then suggests that the Eridanus and NGC 1332 groups are falling into
the NGC 1407 group.  Which may explain why we do not find the Eridanus
and NGC 1332 groups to be bound to one another.  We therefore conclude
that these groups do form a supergroup, that will merge in the future
by dynamical friction to form a cluster.

Using the equations above it is also possible to determine whether the
supergroup-structure as a whole is bound to the Fornax cluster.  The
FOF code also provides parameters for Fornax ($N=145,~v=1404\pm43$ km
s$^{-1}$, $M_V=6.8\times10^{13}M_\odot$, centred on
$\alpha=54.27,\delta=-35.45$ J2000).  The parameters and probabilities
for the supergroup--Fornax pair are given in Table~\ref{bound_table}.
In summary, we find that the supergroup is bound to the Fornax cluster
at the 56 per cent level, which is not significantly probable.
Interestingly, given the observed properties of the two systems, only
bound-outgoing solutions are applicable.  We therefore conclude that
the supergroup may be bound to the Fornax cluster, and if so, it is
currently moving away from that system.



Studies of intra-cluster X-ray emission have concluded that
substructures are common features of clusters of galaxies with
$\sim40$ per cent of clusters with redshifts $z<0.2$ showing
substructure in their X-ray distributions \citep{jones99}.  However,
these subclusters already share common, massive, potential wells, as
evidenced by their X-ray emission.  The subcluster systems are,
therefore, significantly more evolved than the supergroup structure
examined here.  Only 1 of the 25 low-richness clusters studied by
\cite{burgett04} shows any similarity with the Eridanus supergroup:
Abell 1750 is made up of four sub-clumps, two of which show extended
intra-clump X-ray emission, at similar separations to those between
the groups observed here.  However, it is at a much higher redshift
($z\sim0.08$) than Eridanus.

The three groups in the Eridanus region form a rare, local example of
a supergroup.  Owing to its proximity to us it can be studied in more
detail than similar structures at higher redshifts.  It is, therefore,
important to examine the properties of its constituent galaxies and
compare them to previous work on larger scales.

\section{Galaxy Properties}

Previous studies examining the effects of environment on galaxy
evolution have shown that luminous, red, early-type galaxies inhabit
the cores of clusters and that the star formation rates of cluster
galaxies are lower compared to the field population
(e.g. \citealt{faber73,oemler74,visvanathan77,dressler80,butcher84}).
This has recently been established to occur at clustercentric radii of
$\sim3$ virial radii, equivalent to projected surface densities of
$\sim1$ galaxy Mpc$^{-2}$
(e.g. \citealt{lewis02,gomez03,balogh042df}).

Studying such a large area of sky means that our galaxy sample
encompasses a wide range of environments, from the dense cluster
environment of the Fornax cluster, through the
Eridanus supergroup to the galaxies in the field around.  This allows
us to examine the effects of a wide range of environments on a sample
of galaxies.

We calculated the local galaxy density of each galaxy as the projected
surface density of the 5 nearest neighbours to that galaxy within $\pm
1000$ km s$^{-1}$ ($\Sigma_5$; \citealt{balogh042df,tanaka04}).  The
effects of edges in the plane of the sky were solved by excluding any
galaxy whose 5th nearest neighbour is further away than the closest
survey boundary, since the density of these galaxies is not correctly
estimated (c.f. \citealt{balogh042df, tanaka04,tanaka05}).  These
galaxies are still used in the local density calculation of other
galaxies.  To solve the effects of our velocity limit, we extended the
data to recession velocities of 3500 km s$^{-1}$ such that our density
calculations are accurate to 2500 km s$^{-1}$.  We find that densities
of 75--250 Mpc$^{-2}$ correspond to the centre of the Fornax cluster
whilst $\Sigma_5\sim100$ Mpc$^{-2}$ corresponds to the density around
the NGC 1407 galaxy, $\Sigma_5\sim50$ Mpc$^{-2}$ to the NGC 1332
galaxy and $\Sigma_5\sim25$ Mpc$^{-2}$ around the NGC 1395 galaxy.

Using the apparent-magnitude limited sample of 185 galaxies with 2MASS
$K$-band magnitudes $m_K<13.1$ mag and accurate density measurements,
of which 183 have total $B$-band magnitudes and 178 have morphological
T-types, we are able to examine the distribution with environment of
these properties.

In order to test the effects of incompleteness in this dataset we also
repeated these analyses for the sample limited to $m_K\leq$11 mag, and
we find no change in the conclusions we present below.

\subsection{Colours}

The colours of galaxies follow a bimodal distribution
(e.g. \citealt{balogh04sdss,baldry04,blanton05}) depending on their
luminosity, and also their environment.  The reddest galaxies have
long been known to have the highest luminosities
(e.g. \citealt{faber73,visvanathan77}) whilst the fraction of red
galaxies is higher in the densest environments
(e.g. \citealt{oemler74,butcher84,kodama01,girardi03}).
Previous analyses of the group environment have found that the colours
of galaxies in groups are redder than those in the field
\citep{girardi03,tovmassian04}.

\begin{figure}
\begin{center}

    \resizebox{20pc}{!}{
     \rotatebox{-90}{
	\includegraphics{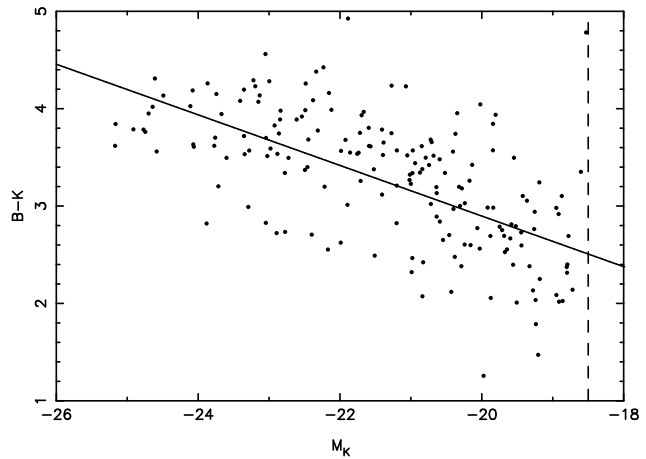}
    }
}
  \end{center}
\caption
{$B-K$ vs $M_K$ (left panel) colour-magnitude diagram.  The solid line
is the straight-line fit to the data given in
Equations~\ref{eq:colour_mag}.  The dashed line indicates the apparent
magnitude limit of these data ($m_K=13.1$ mag).
}
\label{b_kcolour}
\end{figure}

In order to examine the relationship between colour and environment,
we correct the galaxies to a specific mass by correcting the colours
of the galaxies to the colour they would have at a specific $K$
magnitude, $M_{K,spec}$ based on the slope of the colour-magnitude
relation, $m$, i.e. $B-K_c=(B-K)-m(M_K+M_{K,spec})$
(c.f. \citealt{kodama01,tanaka05}).  We therefore fit a
colour-magnitude relation to our data.  We use the non-parametric
IRAF/STSDAS/STATISTICS/Buckley-James algorithm which uses the
Kaplan-Meier estimator for the residuals to calculate the regression.
The best-fit straight line to all 183 apparent-magnitude limited
galaxies, is given by
\begin{equation}
B-K=-0.26~M_K-2.3,
\label{eq:colour_mag}
\end{equation}
with an rms scatter $\sigma=0.59$ mag, and is shown in Figure~\ref{b_kcolour}.  

Comparing with previous studies (e.g. \citealt{baldry04,blanton05}),
it is evident that a strong red sequence is visible above the
fitted colour-magnitude relation and a blue galaxy population is present
below.  The blue galaxy population is smaller than the red sequence, as
would be expected at these magnitudes from the colour functions of
\cite{baldry04}.


\begin{figure}
\begin{center}

    \resizebox{20pc}{!}{
     \rotatebox{-90}{
	\includegraphics{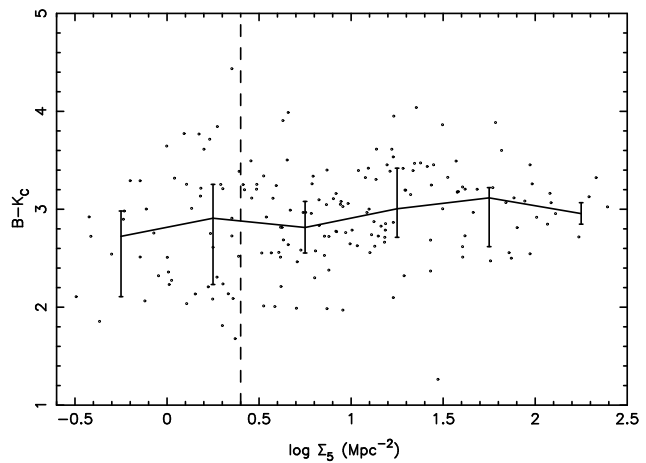}
    }
}
  \end{center}
\caption
{Variation in colour with local projected galaxy density ($\Sigma_5$).
The colours are normalised to $M_K=-20$.  The solid line represents
the loci of the median colours and the error bars indicate the $\pm
25$ percentile colours.  The dashed line indicates the density at
which the colours of the galaxies in the lower and higher density
environments are most different ($2.5$ galaxies Mpc$^{-2}$).
}
\label{colour_dens}
\end{figure}

The colours of the galaxies are corrected to the colour they would
possess at a magnitude of $M_{K,spec}=-20$, assuming the
colour-magnitude relation fitted above.  These corrected colours are
compared with their projected surface density in Figure~\ref{colour_dens}.
A shallow trend of redder colour with increasing galaxy density is
observed.  A non-parametric Spearman rank correlation
gives a correlation coefficient of $r_s=0.25$.  For 183 galaxies, the
Student's t-test rejects the null hypothesis that there is no
correlation at $>99$ per cent confidence level. Thus galaxies in the
densest environments are redder than those in the least dense
environments.

However, the relationship between colour and density does not show a
sharp transition as observed by \cite{kodama01} and \cite{tanaka04},
or by \cite{lewis02} and \cite{gomez03} in galaxy star formation rates
with density.  It is somewhat surprising that we do not observe this
sharp transition as we study the same range of projected surface
densities (after correcting for different values of $H_0$).  However,
the previous studies all had samples at least $\sim 50$ times larger
(e.g. \citealt{gomez03} with 8598 galaxies) than that available here.

We therefore examined the data to determine whether there is a density
at which the colours of the galaxies in denser environments are
significantly different to those in less dense environments.  We
calculated the probability, given by a non-parametric, two-tailed
Kolmogorov-Smirnov (KS) test, that the galaxies in the high- and
low-density subsamples are drawn from the same parent population.  The
minimum in this probability occurs at a projected local surface
density of $\sim2.5$ galaxies Mpc$^{-2}$ with a probability of 99.98
per cent that the high- and low-density samples are not drawn from the
same parent population.  As this probability was determined {\it a
posteriori} we examine the significance of this result using a
Monte-Carlo simulation of the galaxy colours, randomized with respect
to their surface density to remove the correlation with environment.
Repeating this 1000 times, the minimum in the probability distribution
never rose to 99.98 per cent.  We conclude that the colours of the
galaxies in the high- and low-density samples are not drawn from the
same parent population with a probability of 99.98 per cent.


This density is not a break density as there is no sharp transition
in the colours of the galaxies, it marks the density at which the
colours of the galaxies in the lowest density environments are most
different to those in the highest density environments.  A local
projected surface density of $\sim2.5$ galaxies Mpc$^{-2}$ corresponds
to the density of galaxies on the outskirts of the supergroup
structure and other surrounding groups.  It is slightly higher than
the break densities observed by \cite{lewis02,gomez03} and
\cite{balogh042df}, $\Sigma\sim1-2$ galaxies Mpc$^{-2}$ but is
consistent with the break observed by \cite{tanaka04}, $\Sigma\sim2.5$
galaxies Mpc$^{-2}$.  \cite{tanaka04} ascribe the difference in
density to the shallower magnitude cuts used in \cite{lewis02,gomez03}
and \cite{balogh042df}.  We use an equivalent magnitude cut to
\cite{tanaka04} and, taking that into account, this dividing density is
in agreement with previous estimates of break densities.  We therefore
conclude that there are signs of a difference in colour with density
around a density of $\Sigma\sim2.5$ galaxies Mpc$^{-2}$, and that this
value is consistent with previous estimates of break densities,
however a significantly larger sample is required to determine whether
this would translate into a sharp transition.




\subsection{Morphologies and Luminosities}

The morphology-density relation
(e.g. \citealt{dressler80,depropris03}) is one of the best known
segregation effects of galaxies: early-type galaxies are
preferentially found in the densest regions of the Universe.  The
luminosity functions of galaxies are also known to change with their
environment.  \cite{ferguson91,zabludoff00,girardi03,croton05} and
\cite{wilman05} all observe that the luminosity functions of galaxies
in groups have more bright galaxies and less faint galaxies than the
luminosity functions of galaxies in the field.  However,
\cite{croton05} show that the mean luminosity of the galaxies does not
change with environmental density over the range of environments they
study (i.e. void to cluster).

\begin{figure}
\begin{center}

    \resizebox{20pc}{!}{
     \rotatebox{-90}{
	\includegraphics{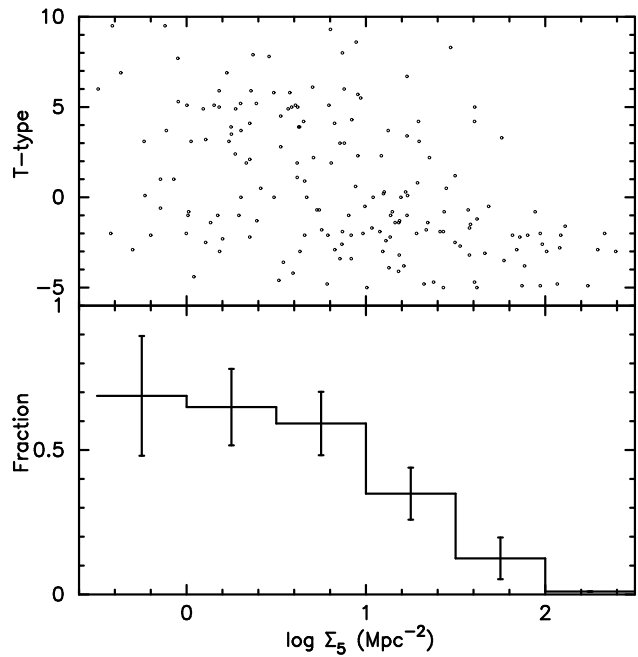}
    }
	
}
  \end{center}
\caption
{The variation in morphological T-types with projected local galaxy
density ($\Sigma_5$; upper panel).  The lower panel shows the fraction
of spiral galaxies (T-types$>0.0$) in each projected local galaxy
density bin, the error bars indicate the poisson error in each bin.}

\label{morph_dens}
\end{figure}

Figure~\ref{morph_dens} compares the morphological distribution of
galaxies with the projected local surface density.  As a scatter plot
of the T-types says more about the divisions used to define the
morphologies, we also indicate the fraction of spiral galaxies
(T-type $>0.0$) in each density bin.
There is a clear dependence of morphology on projected surface density
with galaxies in the least dense environments consisting almost solely
of spiral galaxies, reproducing the morphology-density relationship.

\begin{figure}
\begin{center}

    \resizebox{20pc}{!}{
     \rotatebox{-90}{
	\includegraphics{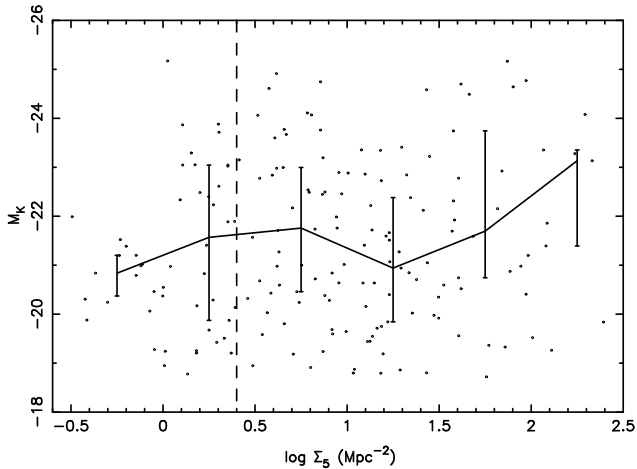}
    }
	
}
  \end{center}
\caption
{Variation in absolute $K$-band magnitude with projected local galaxy
density ($\Sigma_5$).  The solid line represents the loci of the
median magnitudes and the error bars indicate the $\pm 25$ percentile
magnitudes.  The dashed line indicates the density at which the
colours of the galaxies in the lower and higher density environments
are most different ($2.5$ galaxies Mpc$^{-2}$).
}

\label{mag_dens}
\end{figure}

We examine the distribution of the galaxy magnitudes with their
projected surface density in Figure~\ref{mag_dens}.  Whilst we correct
the colours of the galaxies to a standard $K$ magnitude, as a proxy
for a standard mass, in Figure~\ref{colour_dens}, we do not show the
luminosities corrected to a standard colour here.  Adopting this
correction has no effect on the results we obtain.  A shallow trend of
brighter median luminosity with increasing galaxy density is observed.
A non-parametric Spearman rank correlation
gives a correlation coefficient of $r_s=-0.07$.  For 183 galaxies, the
Student's t-test rejects the null hypothesis that there is no
correlation at $>99$ per cent confidence level. Thus galaxies in the
densest environments are brighter than those in the least dense
environments.

\section{Neutral Hydrogen}

\begin{figure}
\begin{center}

    \resizebox{20pc}{!}{
     \rotatebox{-90}{
	\includegraphics{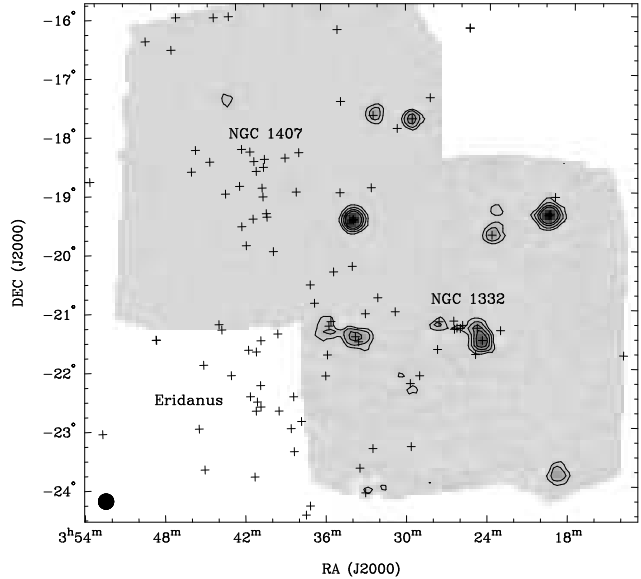}
    }
}
  \end{center}
\caption
{Velocity integrated map of the HI emission ,in the region of the NGC
1332 and NGC 1407 groups from the Parkes radiotelescope.  The contours
indicate HI emission whilst the crosses indicate 6dFGS and NED
sources.  The beam size is indicated in the bottom left-hand corner.}
\label{HI_plot}
\end{figure}

Wide-field neutral hydrogen (HI) observations of the NGC 1407 and NGC
1332 groups were made as part of the GEMS project using the Parkes
radiotelescope (see \citealt{kilborn06} for details of the
observations and data reduction).  A 5.5 $\times$ 5.5 degree region
was mapped around each of the groups, to a sensitivity of $\sim 4
\times 10^8 M_{\odot}$.  Figure~\ref{HI_plot} shows a velocity
integrated map of the HI emission in the region of NGC 1332 and NGC
1407. It is clear that the HI emission is quite different for the two
groups -- the NGC 1407 group shows a total lack of HI in the centre of
the group, while the HI emission is more central for the NGC 1332
group.  The total HI mass within $r_{500}$ for the NGC 1407 group is
just $8\times 10^7 M_{\odot}$, with all the HI contained in one spiral
galaxy, NGC 1390.  This gives a group $M_{HI}/L_B$ ratio of 0.002.
The $r_{500}$ radius for NGC 1332 is much smaller than that of NGC
1407, and the total HI mass contained
is $4.1 \times 10^9 M_{\odot}$, with two galaxies containing the HI,
(NGC 1325 and NGC 1325A).
This gives a group $M_{HI}/L_B$ ratio of 0.12.  The HI content of the
NGC 1332 group is nearly two orders of magnitude greater than the HI
content of the NGC 1407 group.

\cite{omar05a,omar05b} have made HI observations of selected
galaxies in the Eridanus region using the Giant Meterwave Radio
Telescope (GMRT) in India.  If we compare our HI results to that of
\cite{omar05b}, we see that for our overlapping galaxies,
our HI masses are very similar.  Their targeted observations are more
sensitive than GEMS, and they detect three more late-type galaxies in
HI in the NGC 1407 group -- ESO 548- G 065, ESO 548- G 072 and ESO
549- G 002.  All of these galaxies are at group-centric radii $>280$
kpc and are HI deficient with regards to their optical type and
luminosity, with ESO 548- G 072 and ESO 549- G 002 containing about
one quarter as much HI as expected \citep{omar05b}.  They also analyse
the Eridanus group, but because they target a few individual galaxies
for observation rather than scanning the whole area it is not possible
to calculate the total HI mass within that region from their
observations.
\cite{omar05b} postulate that the gas removal mechanism for galaxies in 
the Eridanus region is tidal interactions rather than ram pressure
stripping.

\section{Discussion and Conclusions}

We have defined a supergroup as a 
group of groups that will eventually merge to form a cluster and
examined a possible supergroup in the direction of the Eridanus
constellation.

Our FOF analysis has determined that the region is made up of
three individual groups, with varying properties: The NGC 1407 group
is a massive group with symmetric intra-group X-ray emission centred
on the large central elliptical galaxy, implying that this is a
dynamically mature group.  In contrast, the Eridanus group is a
low-mass, irregular group with a high spiral fraction.  It is not
centred on any one galaxy and the spatial offset of the X-ray emission
signifies that this group is dynamically young.  The NGC 1332 group is
a compact, low-mass group with a low spiral fraction, however there is
no X-ray emission associated with this group, only with the central
galaxy NGC 1332.

Our analysis of the dynamics of these groups indicates that, whilst
the three groups are not gravitationally bound to one another, they
are likely to merge into a single poor cluster, of mass
$M\sim7\times10^{13}M_\odot$.  Furthermore, they may also be bound and
expanding away from the Fornax cluster.  We therefore conclude that
Eridanus is a supergroup.

Previous studies of clusters of galaxies have determined that a large
proportion of clusters show substructure in their X-ray properties
and/or galaxy velocity distributions.  However, this substructure is
usually within a radius of $0.5$ Mpc of the cluster centroid
\citep{burgett04} and the structures share X-ray haloes with their 
host cluster, indicating that they are more compact, more massive and
more tightly bound than the supergroup examined here.  
This supergroup is also a much less massive system than the one
discovered by \cite{gonzalez05}, i.e. SG 1120-1202 with
$M=5.3\times10^{14}M_\odot$.  However, that supergroup is at a
redshift $z\sim0.4$ and is made up of at least four individual groups,
each of which have extended X-ray haloes, indicative of dynamically
mature group structures.

Although one expects clusters to form from the merging of small galaxy
group sized structures, this is expected to happen predominantly at
high redshifts, $z\sim1$.  However, the properties and ratio of masses
of the individual groups studied here argue that the supergroup
consists of one massive, relaxed group (the NGC 1407 group) that
formed at those epochs and now has two less massive groups falling in
to form a cluster-mass structure.  The merging timescale predicted for
this mass of structure indicates that a recession velocity $\sim1600$
km s$^{-1}$ places it at the tail-end of the likely formation times
\citep{lacey93}, but still entirely consistent with the predictions of
hierarchical structure formation in a $\Lambda$CDM Universe.

An examination of the properties of the galaxies in this region
indicates that the morphological T-types show a clear
morphology-density relation.  The three groups studied here have
slightly higher spiral fractions than that of the Fornax cluster,
consistent with their lower projected surface densities.  
The galaxy colours do not show a sharp transition, or break with
density.  However, the distribution of the colours with density shows
the most difference around a projected surface density of
$\Sigma_5\sim2.5$ galaxies Mpc$^{-2}$.  This density is significantly
less than that within the supergroup structure itself and is
equivalent to that of galaxies surrounding the supergroup and in
outlying groups.  This indicates that the colours and luminosities of
the galaxies within the Eridanus supergroup are already similar to
those of the dense core of a cluster like Fornax, whilst their
morphologies show a higher fraction of spiral galaxies, consistent
with their lower densities.

The galaxies within the supergroup are already similar to those in the
Fornax cluster and thus any change is occuring at significantly lower
densities.  If the differences in galaxy properties with environment
{\it are} due to nurture then these observations limit possible
mechanisms for transforming the galaxies from blue, late-types to red,
early-types by their environment to those which are active at low
densities.  We note that SPH/N-body simulations have shown that ram
pressure stripping of cold gas is only effective in the cores of rich
systems, where galaxies are moving at high speeds and there is a
dense, hot, intracluster medium
\citep{abadi99,quilis00}.  Further simulations suggest that tidally
induced collisions of galaxy disk gas clouds should also only be
effective in dense clusters
\citep{byrd90}.  Harrassment, where galaxies in rich systems undergo
high-velocity interactions with other galaxies, is also only effective
in very dense environments (e.g. \citealt{moore96}).  The densities of
groups are not high enough for any of these processes to be
responsible for the relationship of galaxy colour and morphology with
environment.  In contrast, mergers are much more likely in the group
environment as the velocity of the encounter is similar to the orbital
timescale within the galaxy (e.g. \citealt{barnes85}).  Strangulation,
where galaxies lose their halo of gas as they fall in to the larger
halo of a group or cluster leading to a slow decline in SFR as the
galaxy consumes the remaining cold gas, is also more likely in the
group environment (e.g. \citealt{balogh00}).  The detection of HI
deficient galaxies in the NGC 1332 group, where no intra-group X-ray
emission exists, suggests that the gas-removal processes cannot be
entirely due to ram pressure stripping.

This is a rare example of a supergroup in the local Universe.  
The mass ratios and properties of the individual groups are consistent
with the predictions of hierarchical structure formation in a $\Lambda
CDM$ Universe.  The properties of the constituent galaxies indicate
that they are already similar to those of a cluster and that this is
likely to be a result of merging or strangulation processes in
group density environments.

 	




\section*{Acknowledgments}

SB would like to thank Chris Power, Huw Jones, Peder Norberg, Phil
James and Marisa Girardi for helpful discussions.  We would also like
to thank the anonymous referee for their helpful suggestions that have
improved the paper.  This publication makes use of data products from
the Two Micron All Sky Survey (2MASS) which is a joint project of the
University of Massachusetts and the Infrared Processing and Analysis
Center/California Institute of Technology, funded by the National
Aeronautics and Space Administration and the National Science
Foundation.  This research has made use of the NASA/IPAC Extragalactic
Database (NED) which is operated by the Jet Propulsion Laboratory,
California Institute of Technology, under contract with the National
Aeronautics and Space Administration.  It has also made use of the
HyperLEDA database.

\appendix

\section[]{6dFGS Galaxies Data Table}
\label{6df_galaxies}

\begin{table*}
\begin{center}
\caption{Details of the 135 galaxies with recession velocities $500<v<2500$ km s$^{-1}$ from 6dFGS DR2.  
The $K$-band magnitudes are from 2MASS, B-band magnitudes and T-types
are from HyperLEDA.  We assume $m-M=31.6.$ Dashes indicate where this
information is not available.}
\begin{tabular}{clcccccccc}
\hline
6dFGS ID&Alternative Name & RA (J2000)&Dec (J2000)&$v$ (km
s$^{-1}$)&$m_K$&$M_K$&$m_B$&$M_B$&T-type\\
\hline
6dF J0223019-204243& -                              &  2:23: 1.91 & -20:42:43.4 & 1717 & -     &      - &    -  &- &  -\\  
6dF J0226215-241724& -                              &  2:26:21.48 & -24:17:23.7 & 1500 &     - &      - &    -  &- &  -\\  
6dF J0223046-211402& NGC 0908                       &  2:23: 4.55 & -21:14: 1.6 & 1486 &  7.36 &-24.24  & 10.93 &-20.78&   5.2\\  
6dF J0223029-204243& NGC 0907                       &  2:23: 2.94 & -20:42:42.7 & 1672 & 10.14 &-21.47  & 13.34 &-18.40&   7.8\\  
6dF J0226224-241719& UGCA 032                       &  2:26:22.44 & -24:17:18.8 & 1560 & 10.46 &-21.14  & 12.79 &-18.88&   8.2\\  
6dF J0234010-152155& MCG -03-07-035                 &  2:34: 0.99 & -15:21:55.2 & 1461 & - &       -    &  15.29&-16.44&   5.0\\  
6dF J0238140-154833& NGC 1034                       &  2:38:14.02 & -15:48:32.7 & 1467 & 11.73 &-19.88  & 14.55 &-17.19&   9.5\\
\hline 
\end{tabular} 
\end{center}
\end{table*}

\begin{table*}
\begin{center}
\caption{Continued.}
\begin{tabular}{clcccccccc}
\hline
6dFGS ID&Alternative Name & RA (J2000)&Dec (J2000)&$v$ (km
s$^{-1}$)&$m_K$&$M_K$&$m_B$&$M_B$&T-type\\
\hline
6dF J0329007-220848& ESO 548- G 025                 &  3:29: 0.69 & -22: 8:47.7 & 1706 & - &       -    & 14.98 &-16.76&   1.1\\ 
6dF J0327422-214159& ESO 548- G 022                 &  3:27:42.16 & -21:41:58.6 & 1295 & - &       -    & 15.64 &-16.09&   5.0\\  
6dF J0333277-213353& ESO 548- G 036                 &  3:33:27.69 & -21:33:52.9 & 1520 & 10.54 &-21.07  & 14.89 &-16.84&   0.1\\  
6dF J0327356-211341& ESO 548- G 021                 &  3:27:35.57 & -21:13:41.4 & 1745 & -     & -      & 14.67 &-17.07&   7.7\\  
6dF J0332576-210522& ESO 548- G 034                 &  3:32:57.63 & -21: 5:21.9 & 1707 & 12.05 &-19.56  & 14.56 &-17.16&   5.0\\  
6dF J0332030-204908& NGC 1353                       &  3:32: 2.98 & -20:49: 8.2 & 1587 &  8.20 &-23.41  & 12.41 &-19.33&   3.1\\  
6dF J0335146-202225& NGC 1370                       &  3:35:14.57 & -20:22:25.2 & 1070 & 10.02 &-21.60  & 13.93 &-17.79&  -3.8\\
\hline 
\end{tabular} 
\end{center}
\end{table*}

\begin{table*}
\begin{center}
\caption{Continued.}
\begin{tabular}{clcccccccc}
\hline
6dFGS ID&Alternative Name & RA (J2000)&Dec (J2000)&$v$ (km
s$^{-1}$)&$m_K$&$M_K$&$m_B$&$M_B$&T-type\\
\hline
6dF J0352420-184518& 2MASX J03524198-1845179        &  3:52:41.97 & -18:45:17.8 & 1880 & 12.67 &-18.95  & 15.88 &-15.97&   0.0\\  
6dF J0349534-321533& FCCB 2144                      &  3:49:53.36 & -32:15:32.7 & 1202 & - &    -       & -     &   -  &  -\\  
6dF J0400416-304952& ESO 419- G 013                 &  4: 0:41.61 & -30:49:51.6 & 1515 & 12.36 &-19.24  & 14.43 &-17.21&  -0.8\\  
6dF J0355228-280930& IC 2007                        &  3:55:22.76 & -28: 9:30.0 & 1292 & 10.59 &-21.02  & 13.90 &-17.75&   3.7\\  
6dF J0408076-171134& NGC 1519                       &  4: 8: 7.60 & -17:11:34.4 & 1810 & 10.41 &-21.20  & 13.72 &-17.99&   3.1\\  
6dF J0357287-231048& -                              &  3:57:28.68 & -23:10:48.1 & 1579 & - &    -       & -     & -    &  -\\  
6dF J0406497-211022& NGC 1518                       &  4: 6:49.72 & -21:10:21.5 &  710 & 11.16 &-20.46  & 12.26 &-19.55&   7.7\\  
6dF J0412411-230932& ESO 483- G 013                 &  4:12:41.12 & -23: 9:32.0 &  905 & 11.38 &-20.24  & 14.19 &-17.64&  -2.9\\  
6dF J0421137-215045& UGCA 090                       &  4:21:13.73 & -21:50:45.3 &  858 & 10.77 &-20.84  & 12.94 &-18.76&   6.9\\  
6dF J0427198-223333& ESO 484- G 019                 &  4:27:19.83 & -22:33:33.3 & 1611 & - &       -    & 16.05 &-15.68&   5.5\\  \hline       
 \hline       

\end{tabular} 
\end{center}
\end{table*}

\section[]{NED Galaxies Data Table}
\label{ned_galaxies}

\begin{table*}
\begin{center}
\caption{Details of the 378 galaxies with recession velocities 
$500<v<2500$ km s$^{-1}$ from the NED database.  The $K$-band
magnitudes are from 2MASS, B-band magnitudes and T-types are from
HyperLEDA.  We assume $m-M=31.6.$ Dashes indicate where information is
not available.}
\begin{tabular}{lcccccccc}
\hline
Galaxy Name & RA (J2000)&Dec (J2000)&$v$ (km s$^{-1}$)&$m_K$&$M_K$&$m_B$&$M_B$&T-type\\
\hline
ESO 545- G 005                 &  2:20: 6.11 & -19:45: 2.0 & 2338 & 10.84 & -21.77 & 13.61 & -18.71 &   8.6\\  
UGCA 027                       &  2:22: 7.08 & -21: 7:54.0 & 1622 & - &          - & 15.86 & -15.83 &  10.0\\  
ESO 415- G 023                 &  2:22:29.32 & -28:51:20.0 & 1615 & - &          - & 16.32 & -15.35 &   7.5\\  
MCG -03-07-009                 &  2:24:14.73 & -19:49:30.0 & 1052 & - &          - & 15.52 & -16.22 &   5.5\\  
2dFGRS S230Z178                &  2:24:35.71 & -26:48:32.0 & 1096 & - &          - & - &          - &  -\\  
ESO 545- G 015                 &  2:25:57.94 & -19:41:29.0 & 2286 & - &          - & 16.72 & -15.03 &  10.0\\  
VV 525                         &  2:26:21.30 &  -9:50:27.0 & 2109 & - &          - & 14.54 & -17.17 &   8.0\\
\hline
\end{tabular} 
\end{center}
\end{table*}

\begin{table*}
\begin{center}
\caption{Continued.}
\begin{tabular}{lcccccccc}
\hline
Galaxy Name & RA (J2000)&Dec (J2000)&$v$ (km s$^{-1}$)&$m_K$&$M_K$&$m_B$&$M_B$&T-type\\
\hline
SHOC 138a                      &  2:49: 9.32 &  -7:50:27.0 & 1290 & - &          - & - &          - &  -\\  

SDSS J024911.16-082828.7       &  2:49:11.16 &  -8:28:28.0 & 1461 & - &          - & - &          - &  -\\  
SDSS J024913.41-080653.2       &  2:49:13.42 &  -8: 6:53.0 & 1369 & - &          - & - &          - &  -\\  
SDSS J024911.16-082828.7       &  2:49:11.16 &  -8:28:28.0 & 1461 & - &          - & - &          - &  -\\  
SDSS J024913.41-080653.2       &  2:49:13.42 &  -8: 6:53.0 & 1369 & - &          - & - &          - &  -\\  
LSBG F416-012                  &  2:51:47.97 & -30: 6:30.0 & 1076 & - &          - & - &          - &  -\\  
SDSS J025328.64-085905.5       &  2:53:28.64 &  -8:59: 5.0 & 1457 & - &          - & - &          - &  -\\
\hline
\end{tabular} 
\end{center}
\end{table*}

\begin{table*}
\begin{center}
\caption{Continued.}
\begin{tabular}{lcccccccc}
\hline
Galaxy Name & RA (J2000)&Dec (J2000)&$v$ (km s$^{-1}$)&$m_K$&$M_K$&$m_B$&$M_B$&T-type\\
\hline
FCSS J033805.0-352409          &  3:38: 5.08 & -35:24: 9.0 & 1244 & - &          - & - &          - &  -\\  
FCSS J033806.3-352858          &  3:38: 6.29 & -35:28:58.0 & 1312 & - &          - & - &          - &  -\\  
FCOS 2-2153                    &  3:38: 6.48 & -35:23: 3.0 & 1390 & - &          - & - &          - &  -\\  
FCOS 2-2134                    &  3:38:10.73 & -35:25:46.0 & 1683 & - &          - & - &          - &  -\\  
FCOS 2-2127                    &  3:38:11.69 & -35:27:16.0 & 1476 & - &          - & - &          - &  -\\  
FCOS 0-2023                    &  3:38:12.70 & -35:28:57.0 & 1705 & - &          - & - &          - &  -\\  
CGF 03-29                      &  3:38:14.75 & -35:33:39.0 & 1331 & - &          - & - &          - &  -\\
\hline
\end{tabular} 
\end{center}
\end{table*}

\begin{table*}
\begin{center}
\caption{Continued.}
\begin{tabular}{lcccccccc}
\hline
Galaxy Name & RA (J2000)&Dec (J2000)&$v$ (km s$^{-1}$)&$m_K$&$M_K$&$m_B$&$M_B$&T-type\\
\hline
FCC 278                        &  3:42:27.10 & -33:52:15.0 & 2125 & - &          - & 16.25 & -15.37 &  -5.0\\  
ESO 548- G 082                 &  3:42:43.27 & -17:30:26.0 & 1716 & - &  -         & 15.38 & -16.69 &   5.0\\  
NGC 1436                       &  3:43:37.08 & -35:51:10.0 & 1387 &  9.02 & -22.58 & 12.86 & -18.78 &   3.3\\  
ESO 482- G 039                 &  3:43:57.20 & -24:55:21.0 & 1381 & - &          - & 17.64 & -14.01 &  10.0\\  
ESO 549- G 007                 &  3:44:11.45 & -19:19:12.0 & 1527 & - &          - & 15.45 & -16.39 &  -1.4\\  
FCC 298                        &  3:44:44.40 & -35:41: 1.0 & 1719 & - &          - & 16.10 & -15.56 &  -5.0\\  
ESO 358- G 059                 &  3:45: 3.56 & -35:58:21.0 & 1007 & 10.79 & -20.81 & 14.12 & -17.52 &  -3.2\\
\hline
\end{tabular} 
\end{center}
\end{table*}

\begin{table*}
\begin{center}
\caption{Continued.}
\begin{tabular}{lcccccccc}
\hline
Galaxy Name & RA (J2000)&Dec (J2000)&$v$ (km s$^{-1}$)&$m_K$&$M_K$&$m_B$&$M_B$&T-type\\
\hline
ESO 357- G 012                 &  3:16:53.35 & -35:32:28.7 & 1567 & - &       -    & 14.78 & -16.92 &   7.0\\  
NGC 1300                       &  3:19:41.06 & -19:24:40.3 & 1577 &  7.90 &-23.72  & 11.22 & -20.51 &   3.7\\  
NGC 1302                       &  3:19:51.21 & -26: 3:38.1 & 1710 &  7.94 &-23.67  & 11.98 & -19.73 &   0.0\\  
ESO 481- G 021                 &  3:20:17.71 & -26:27:50.3 & 1753 & - &       -    & 16.34 & -15.33 &   6.8\\  
2MASX J03202693-3114531        &  3:20:26.85 & -31:14:52.6 & 1830 & 13.96 &-17.65  & 16.11 & -15.55 &  -\\  
NGC 1309                       &  3:22: 6.56 & -15:24: 0.2 & 2136 &  9.22 &-22.40  & 12.08 & -19.69 &   3.9\\  
ESO 357- G 025                 &  3:23:37.26 & -35:46:42.2 & 1823 & 12.98 &-18.63  & 15.18 & -16.49 &  -2.7\\
\hline
\end{tabular} 
\end{center}
\end{table*}

\begin{table*}
\begin{center}
\caption{Continued.}
\begin{tabular}{lcccccccc}
\hline
Galaxy Name & RA (J2000)&Dec (J2000)&$v$ (km s$^{-1}$)&$m_K$&$M_K$&$m_B$&$M_B$&T-type\\
\hline

NGC 1390                       &  3:37:52.17 & -19: 0:30.1 & 1207 & 11.70 &-19.92  & 14.94 & -16.93 &   1.2\\  
NGC 1396                       &  3:38: 6.59 & -35:26:23.7 &  808 & 12.08 &-19.52  & 14.93 & -16.73 &  -3.0\\  
ESO 358- G 042                 &  3:38: 9.17 & -34:31: 7.2 & 1138 & 13.32 &-18.29  & 15.49 & -16.16 &  -1.9\\  
ESO 482- G 018                 &  3:38:17.64 & -23:25: 9.0 & 1687 & 11.77 &-19.84  & 15.43 & -16.27 &   0.3\\  
NGC 1399                       &  3:38:29.01 & -35:27: 1.7 & 1425 &  6.44 &-25.16  & 10.33 & -21.32 &  -4.9\\  
NGC 1395                       &  3:38:29.72 & -23: 1:38.7 & 1717 &  7.02 &-24.58  & 10.68 & -21.02 &  -5.0\\  
NGC 1393                       &  3:38:38.58 & -18:25:40.7 & 2127 &  9.31 &-22.31  & 13.36 & -18.54 &  -1.7\\
\hline
\end{tabular} 
\end{center}
\end{table*}

\begin{table*}
\begin{center}
\caption{Continued.}
\begin{tabular}{lcccccccc}
\hline
Galaxy Name & RA (J2000)&Dec (J2000)&$v$ (km s$^{-1}$)&$m_K$&$M_K$&$m_B$&$M_B$&T-type\\
\hline

DUKST 481-041                  &  3:25:53.58 & -25:48:39.5 & 1828 & - &       -    & 16.66 & -15.00 &  -\\  
APMUKS(BJ) B033830.70-222643.7 &  3:40:41.35 & -22:17:10.5 & 1737 & - &       -    & - &     -      &  -\\  
2MASX J03184089-2545279        &  3:18:40.87 & -25:45:27.8 & 1678 & 13.85 &-17.76  & 16.25 & -15.40 &   0.0\\  
2MASX J04125330-3118316        &  4:12:53.33 & -31:18:29.5 & 1391 & 13.39 &-18.22  & 15.51 & -16.19 &   4.0\\  
2MASX J04110053-3124413        &  4:11: 0.49 & -31:24:41.1 & 1367 & - &       -    & - &     -	    &  -\\  
2MASX J03481580-2128131        &  3:48:15.81 & -21:28:13.2 & 1587 & 13.21 &-18.41  & - &     -	    &  -\\  
APMUKS(BJ) B041228.64-225548.4 &  4:14:37.41 & -22:48:25.1 & 2278 & - &       -    & - &     -	    &  -\\  
APMBGC 550+098+084             &  4: 6:18.37 & -21:31:30.4 & 1010 & - &       -    & - &     -	    &  -\\  
LSBG F358-091                  &  3:32:10.73 & -36:13:13.7 & 1199 & - &    -       & - &     -	    &  -\\  
2MASX J03231011-3437594        &  3:23:10.12 & -34:37:59.4 & 1522 & 13.07 &-18.53  & 17.90 & -13.75 &  -\\  
FCCB 0692                      &  3:31: 0.19 & -35:28: 9.1 & 1795 & 13.19 &-18.42  & - &     -      &  -\\  
ESO 548- G 046                 &  3:34:38.12 & -17:28:34.8 & 1567 & - &    -       & 15.69 & -16.24 &   3.9\\ \hline
\hline
\end{tabular} 
\end{center}
\end{table*}

\bsp

\label{lastpage}


\begin{thebibliography}{5}

\bibitem[\protect\citeauthoryear{Abadi et al.}{1999}]{abadi99}
{Abadi~M.~G., Moore~B., Bower~R.~G., 1999, MNRAS, 308, 947}


\bibitem[\protect\citeauthoryear{Abazajian et al.}{2003}]{abazajian03}
{Abazajian~K. et al. 2003, AJ, 126, 2081}

\bibitem[\protect\citeauthoryear{Baker \& Shapley}{1933}]{baker33}
{Baker~R.~H., Shapley~H., 1933, Annals of the Astronomical Observatory
of Harvard College, 88, 77}

\bibitem[\protect\citeauthoryear{Baldry et al.}{2004}]{baldry04}
{Baldry~I.~K., Glazebrook~K., Brinkmann~J., Ivezi\'{c} $\rm{\check{Z}}$.,  Lupton~R.~H., Nichol~R.~C., Szalay~A.~S., 2004, ApJ, 600, 681}

\bibitem[\protect\citeauthoryear{Balogh et al.}{1997}]{balogh97}
{Balogh~M.~L., Morris~S.~L., Yee~H.~K.~C., Carlberg~R.~G., Ellingson~E., 1997, ApJ, 488, 75}

\bibitem[\protect\citeauthoryear{Balogh et al.}{1998}]{balogh98}
{Balogh~M.~L., Schade~D., Morris~S.~L., Yee~H.~K.~C., Carlberg~R.~G. \& Ellingson~E., 1998, ApJ, 504, L75}

\bibitem[\protect\citeauthoryear{Balogh et al.}{2000}]{balogh00}
{Balogh~M.~L., Navarro~J.~F., Morris~S.~L., 2000, ApJ, 540, 113}

\bibitem[\protect\citeauthoryear{Balogh et al.}{2004}]{balogh04sdss}
{Balogh~M.~L., Baldry~I.~K., Nichol~R., Miller~C., Bower~R., Glazebrook~K., 2004, ApJ, 615, 101}

\bibitem[\protect\citeauthoryear{Balogh et al.}{2004}]{balogh042df}
{Balogh~M.~L. et al., 2004, MNRAS, 348, 1355}

\bibitem[\protect\citeauthoryear{Barnes}{1985}]{barnes85}
{Barnes~J., 1985, MNRAS, 215, 517}

\bibitem[\protect\citeauthoryear{Barton et al.}{1996}]{barton96}	
{Barton~E., Geller~M., Ramella~M., Marzke~R.~O., da Costa~L.~N., 1996, AJ, 112,87}

\bibitem[\protect\citeauthoryear{Beers et al.}{1982}]{beers82}
{Beers~T.~C., Geller~M.~J., Huchra~J.~P., 1982, ApJ, 257, 23}

\bibitem[\protect\citeauthoryear{Beers et al.}{1990}]{beers90}
{Beers~T.~C., Flynn~K., Gebhardt~K., 1990, AJ, 100, 32}

\bibitem[\protect\citeauthoryear{Biviano et al.}{1993}]{biviano93}
{Biviano~A., Girardi~M., Giuricin~G., Mardirossian~F., Mezzetti~M., 1993, ApJ, 411, 13}

\bibitem[\protect\citeauthoryear{Blanton et al.}{2003}]{blanton03}
{Blanton~M.~R. et al., 2003, ApJ, 594, 186}


\bibitem[\protect\citeauthoryear{Blanton et al.}{2005}]{blanton05}
{Blanton~M.~R., Eisenstein~D., Hogg~D.~W., Schlegel~D.~J., Brinkmann~J., 2005, ApJ, 629, 143}

\bibitem[\protect\citeauthoryear{Blumenthal et al.}{1984}]{blumenthal84}
{Blumenthal~G.~R., Faber~S.~M., Primack~J.~R., Rees~M.~J., 1984, Nature, 311, 517}

\bibitem[\protect\citeauthoryear{Burgett et al.}{2004}]{burgett04}
{Burgett~W.~S., 2004, MNRAS, 352, 605}

\bibitem[\protect\citeauthoryear{Butcher \& Oemler}{1984}]{butcher84}
{Butcher~H., Oemler~A.~Jr., 1984, ApJ, 285, 426}

\bibitem[\protect\citeauthoryear{Byrd \& Valtonen}{1990}]{byrd90}
{Byrd~G., Valtonen~M., 1990, ApJ, 350, 89}

\bibitem[\protect\citeauthoryear{Cole et al.}{2000}]{cole00}
{Cole~S., Lacey~C.~G., Baugh~C.~M., Frenk~C.~S., 2000, MNRAS, 319, 168}

\bibitem[\protect\citeauthoryear{Cortese et al.}{2004}]{cortese04}
{Cortese~L., Gavazzi~G., Boselli~A., Iglesias-Paramo~J., Carrasco~L., 2004, A\&A, 425, 429}

\bibitem[\protect\citeauthoryear{Colless et al.}{2001}]{colless01}
{Colless~M. et al., 2001, MNRAS, 328, 1039}

\bibitem[\protect\citeauthoryear{Croton et al}{2005}]{croton05}
{Croton~D.~J., 2005, MNRAS, 356, 1155}

\bibitem[\protect\citeauthoryear{da Costa et al.}{1988}]{dacosta88}
{da Costa,~L.~N. et al., 1988, ApJ, 327, 544}

\bibitem[\protect\citeauthoryear{da Costa et al.}{1998}]{dacosta98}
{da Costa,~L.~N. et al., 1998, AJ, 116, 1}

\bibitem[\protect\citeauthoryear{de Propris et al.}{2003}]{depropris03}
{de Propris R. et al., 2003, MNRAS, 342, 725}

\bibitem[\protect\citeauthoryear{de Vaucouleurs}{1975}]{devaucouleurs75}
{de Vaucouleurs~G., 1975, Stars and stellar systems, 9, 557}

\bibitem[\protect\citeauthoryear{Diaferio et al.}{1999}]{diaferio99}
{Diaferio~A., Kauffmann~G., Colberg~J.~M., White~S.~D.~M., 1999, MNRAS, 307, 537}

\bibitem[\protect\citeauthoryear{Dressler}{1980}]{dressler80}
{Dressler~A., 1980, ApJ, 236, 351}

\bibitem[\protect\citeauthoryear{Dressler et al.}{1997}]{dressler97}
{Dressler~A. et al., 1997, ApJ, 490, 577}


\bibitem[\protect\citeauthoryear{Eke et al.}{2004}]{eke04}
{Eke~V.~R. et al., 2004, MNRAS, 348, 866}

\bibitem[\protect\citeauthoryear{Faber}{1973}]{faber73}
{Faber S. M., 1973, ApJ, 179, 731}

\bibitem[\protect\citeauthoryear{Ferguson \& Sandage}{1991}]{ferguson91}
{Ferguson~H.~C., Sandage~A., 1991, AJ, 101, 765}

\bibitem[\protect\citeauthoryear{Forbes et al.}{2005}]{forbes05}
{Forbes~D.~A. et al. 2005, submitted to MNRAS}

\bibitem[\protect\citeauthoryear{Garcia}{1993}]{garcia93}
{Garcia~A.~M., 1993, A\&AS, 100, 47}

\bibitem[\protect\citeauthoryear{Girardi et al.}{2003}]{girardi03}
{Girardi~M., Mardirossian~F., Marinoni~C., Mezzetti~M. \& Rigoni~E., 2003, A\&A, 410, 461}

\bibitem[\protect\citeauthoryear{Girardi et al.}{2005}]{girardi05}
{Girardi~M., Demarco~R., Rosati~P., Borgani~S. 2005, A\&A, 442, 29}

\bibitem[\protect\citeauthoryear{Giuricin et al.}{2000}]{giuricin00}
{Giuricin~G., Marinoni~C., Ceriani~L., Pisani~A., 2000, ApJ, 543, 178}

\bibitem[\protect\citeauthoryear{Gnedin}{2003}]{gnedin03}
{Gnedin~O.~Y. 2003, ApJ, 582, 141}

\bibitem[\protect\citeauthoryear{G\'{o}mez et al.}{2003}]{gomez03}
{G\'{o}mez~P.~L. et al., 2003, ApJ, 584, 210}


\bibitem[\protect\citeauthoryear{Gonzalez et al.}{2005}]{gonzalez05}
{Gonzalez~A.~H., Tran~K.~H., Conbere~M.~N. \& Zaritsky~D., 2005, ApJ, 624, 73}

\bibitem[\protect\citeauthoryear{Gould}{1993}]{gould93}
{Gould~A., 1993, ApJ, 403, 37}

\bibitem[\protect\citeauthoryear{Gregory \& Thompson}{1984}]{gregory84}
{Gregory~S. A., Thompson~L., ApJ, 286, 422}

\bibitem[\protect\citeauthoryear{Gunn \& Gott}{1972}]{gunn72}
{Gunn~J.~E., Gott~J~R., 1972, ApJ, 176, 1}

\bibitem[\protect\citeauthoryear{Hambly et al.}{2001}]{hambly01}
{Hambly~N.~C. et al., 2001, MNRAS, 326, 1279}


\bibitem[\protect\citeauthoryear{Hashimoto et al.}{1998}]{hashimoto98}
{Hashimoto~Y., Oemler~A.~Jr., Lin~H., Tucker~D.~L., 1998, ApJ, 499, 589}

\bibitem[\protect\citeauthoryear{Hashimoto \& Oemler}{2000}]{hashimoto00}
{Hashimoto~Y., Oemler~A.~Jr., 2000, ApJ, 530, 652}

\bibitem[\protect\citeauthoryear{Heisler et al.}{1985}]{heisler85}
{Heisler, Tremaine, Bahcall 1985, ApJ, 298, 8}

\bibitem[\protect\citeauthoryear{Helsdon \& Ponman}{2000}]{helsdon00}
{Helsdon~S.~F., Ponman~T.~J., 2000, MNRAS, 315, 356}

\bibitem[\protect\citeauthoryear{Hilton et al.}{2005}]{hilton05}
{Hilton~M., 2005, MNRAS, accepted}

\bibitem[\protect\citeauthoryear{Huchra \& Geller}{1982}]{huchra82}
{Huchra~J.~P., Geller~M.~J., 1982, ApJ, 257, 423}


\bibitem[\protect\citeauthoryear{Jarrett et al.}{2000}]{jarrett00}
{Jarrett~T.~H., Chester~T., Cutri~R., Schneider~S., Skrutskie~M. \& Huchra~J.~P., 2000, AJ, 119, 2498}

\bibitem[\protect\citeauthoryear{Jensen et al.}{2003}]{jensen03}
{Jensen~J.~B. et al., 2003, ApJ, 583, 712}

\bibitem[\protect\citeauthoryear{Jones \& Forman}{1999}]{jones99}
{Jones~C., Forman~W., 1999, ApJ, 511, 65}

\bibitem[\protect\citeauthoryear{Jones et al.}{2004}]{jones04}
{Jones~D.~H. et al., 2004, MNRAS, 355, 747}


\bibitem[\protect\citeauthoryear{Jones et al.}{2005}]{jones05}
{Jones~D.~H., Saunders~W., Read~M., Colless~M., 2005, astro-ph/0505068}

\bibitem[\protect\citeauthoryear{Kauffmann et al.}{2004}]{kauffmann04}
{Kauffmann~G. et al., 2004, MNRAS, 353, 713}

\bibitem[\protect\citeauthoryear{Kilborn et al.}{2005}]{kilborn05}
{Kilborn~V.~A. Koribalski~B.~S., Forbes~D.~A., Barnes~D.~G., Musgrave~R.~C., 2005, MNRAS, 356, 77}

\bibitem[\protect\citeauthoryear{Kilborn et al.}{2006}]{kilborn06}
{Kilborn~V.~A. et al., 2006, in preparation}

\bibitem[\protect\citeauthoryear{Kochanek et al.}{2001}]{kochanek01}
{Kochanek~C.~S. et al., 2001, ApJ, 560, 566}

\bibitem[{{Kodama} {et~al.}(2001){Kodama}, {Smail}, {Nakata}, {Okamura}, \& {Bower}}]{kodama01}{Kodama}, T., {Smail}, I., {Nakata}, F., {Okamura}, S., {Bower}, R.~G. 2001, ApJL, 562, 9

\bibitem[\protect\citeauthoryear{Lacey \& Cole}{1993}]{lacey93}
{Lacey~C.~G., Cole~S., 1993, MNRAS, 262, 627}

\bibitem[\protect\citeauthoryear{Lares et al.}{2004}]{lares04}
{Lares~M., Lambas~D.~G., S\'{a}nchez~A.~G., 2004, MNRAS, 352, 501}

\bibitem[\protect\citeauthoryear{Larson et al.}{1980}]{larson80}
{Larson~R.~B., Tinsley~B.~M., Caldwell~C.~N., 1980, ApJ, 237, 692}

\bibitem[\protect\citeauthoryear{Lewis et al.}{2002}]{lewis02}
{Lewis~I. et al., 2002, MNRAS, 334, 673}


\bibitem[\protect\citeauthoryear{Marinoni et al.}{1998}]{marinoni98}
{Marinoni~C., Monaco~P., Giuricin~G., Costantini~B., 1998, ApJ, 505, 484}

\bibitem[\protect\citeauthoryear{Mieske et al.}{2004}]{mieske04}
{Mieske~S., Hilker~M., Infante~L., 2004, A\&A, 418, 445}

\bibitem[\protect\citeauthoryear{Miles et al.}{2004}]{miles04}
{Miles~T.~A., Raychaudhury~S., Forbes~D.~A., Goudfrooij~P., Ponman~T.~J., Kozhurina-Platais~V., 2004, MNRAS, 355, 785}

\bibitem[\protect\citeauthoryear{Moore et al.}{1996}]{moore96}
{Moore~B., Katz~N., Lake~G., Dressler~A., Oemler~A., 1996, Nature, 379, 613}
	
\bibitem[\protect\citeauthoryear{Nolthenius \& White}{1987}]{nolthenius87}
{Nolthenius~R., White~S.~D.~M., 1987, MNRAS, 225, 505}

\bibitem[\protect\citeauthoryear{Nolthenius et al.}{1997}]{nolthenius97}
{Nolthenius~R., Klypin~A.~A., Primack~J.~R., 1997, ApJ, 480, 43}

\bibitem[\protect\citeauthoryear{Oemler}{1974}]{oemler74}
{Oemler~A.~Jr., 1974, ApJ, 194, 1O}

\bibitem[\protect\citeauthoryear{Omar \& Dwarakanath}{2005a}]{omar05a}
{Omar~A. \& Dwarakanath~K.~S., 2005, JApA, 26, 1}

\bibitem[\protect\citeauthoryear{Omar \& Dwarakanath}{2005b}]{omar05b}
{Omar~A. \& Dwarakanath~K.~S., 2005, JApA, 26, 71}

\bibitem[\protect\citeauthoryear{Osmond \& Ponman}{2004}]{osmond04}
{Osmond~J.~P.~F., Ponman~T.~J., 2004, MNRAS, 350, 1511}

\bibitem[\protect\citeauthoryear{Paturel et al.}{1997}]{paturel97}
{Paturel~G. et al., 1997, A\&AS, 124, 109}


\bibitem[\protect\citeauthoryear{Perrett et al.}{1997}]{perrett97}
{Perrett~K.~M., Hanes~D.~A., Butterworth~S.~T., Kavelaars~Jj, Geisler~D., Harris~W.~E., 1997, AJ, 113, 895}

\bibitem[\protect\citeauthoryear{Pimbblet et al.}{2002}]{pimbblet02}
{Pimbblet~K.~A., Smail~I., Kodama~T., Couch~W.~J., Edge~A.~C., Zabludoff~A.~I., O'Hely~E., 2002, MNRAS, 331, 333}

\bibitem[\protect\citeauthoryear{Ponman et al.}{1996}]{ponman96} 
{Ponman~T.~J., Bourner~P.~D.~J., Ebeling~H., Bohringer~H., 1996,
MNRAS, 283, 690}

\bibitem[\protect\citeauthoryear{Postman \& Geller}{1984}]{postman84}
{Postman~M., Geller~M.~J., 1984, ApJ, 281 ,95}

\bibitem[\protect\citeauthoryear{Quilis et al.}{2000}]{quilis00}
{Quilis~V., Moore~B., Bower~R., 2000, Science, 288, 1617}

\bibitem[\protect\citeauthoryear{Quintana et al.}{1994}]{quintana94}
{Quintana~H., Fouque~P. \& Way~M.~J., 1994, A\&A, 283, 722}

\bibitem[\protect\citeauthoryear{Ramella et al.}{1995}]{ramella95}
{Ramella~M., Geller~M.~J., Huchra~J.~P., Thorstensen~J.~R., 1995, AJ, 109, 1458}

\bibitem[\protect\citeauthoryear{Ramella et al.}{2002}]{ramella02}
{Ramella~M., Geller~M.~J., Pisani~A., da Costa~L.~N., 2002, AJ, 123, 2976}

\bibitem[\protect\citeauthoryear{Ramella et al.}{2004}]{ramella04}
{Ramella~M., Boschin~W., Geller~M.~J., Mahdavi~A., Rines~K., 2004, AJ,
128, 2022}

\bibitem[\protect\citeauthoryear{Schlegel et al.}{1998}]{schlegel98}
{Schlegel~D.~J., Finkbeiner~D.~P., Davis~M., 1998, ApJ, 500, 525}

\bibitem[\protect\citeauthoryear{Tanaka et al.}{2004}]{tanaka04}
{Tanaka~M., Goto~T., Okamura~S., Shimasaku~K., Brinkmann~J., 2004, AJ, 128, 2677}

\bibitem[\protect\citeauthoryear{Tanaka et al.}{2005}]{tanaka05}
{Tanaka~M. et al., 2005, MNRAS, 362, 268}


\bibitem[\protect\citeauthoryear{Tonry et al.}{2001}]{tonry01}
{Tonry~J.~L. et al., 2001, ApJ, 546, 681}

\bibitem[\protect\citeauthoryear{Tovmassian et al.}{2004}]{tovmassian04}
{Tovmassian~H.~M., Plionis~M., Andernach~H., 2004, ApJ, 617, 111}

\bibitem[\protect\citeauthoryear{Tran et al.}{2001}]{tran01}
{Tran~K-V.~H., Simard~L., Zabludoff~A.~I., Mulchaey~J.~S., 2001, ApJ, 549, 172}

\bibitem[\protect\citeauthoryear{Visvanathan \& Sandage}{1977}]{visvanathan77}
{Visvanathan N., Sandage A,  1977, ApJ, 216, 214}

\bibitem[\protect\citeauthoryear{Wake et al.}{2005}]{wake05}
{Wake~D.~A., Collins~C.~A., Nichol~R.~C., Jones~L.~R., Burke~D.~J., 2005, ApJ, 627, 186}

\bibitem[\protect\citeauthoryear{Wilman et al.}{2005}]{wilman05}
{Wilman~D.~J. et al., 2005, MNRAS, 358, 71}


\bibitem[\protect\citeauthoryear{Willmer et al.}{1989}]{willmer89}
{Willmer~C.~N.~A., Focardi~P., da Costa~L.~N. \& Pellegrini~P.~S., 1989, AJ, 98, 1531}

\bibitem[\protect\citeauthoryear{Zabludoff \& Mulchaey}{1998}]{zabludoff98}
{Zabludoff~A.~I., Mulchaey~J.~S., 1998, ApJ, 496, 39}

\bibitem[\protect\citeauthoryear{Zabludoff \& Mulchaey}{2000}]{zabludoff00}
{Zabludoff~A.~I., Mulchaey~J.~S., 2000, ApJ, 539, 136}

\end{thebibliography}
\end{document}